# Digital filters with vanishing moments for shape analysis

**Hugh L. Kennedy**[*]
DST Group, NSID, Edinburgh, Adelaide, Australia, SA 5111

**Abstract**. Shape- and scale-selective digital-filters, with steerable finite/infinite impulse responses (FIR/IIRs) and non-recursive/recursive realizations, that are separable in both spatial dimensions and adequately isotropic, are derived. The filters are conveniently designed in the frequency domain via derivative constraints at dc, which guarantees orthogonality and monomial selectivity in the pixel domain (i.e. vanishing moments), unlike more commonly used FIR filters derived from Gaussian functions. A two-stage low-pass/high-pass architecture, for blur/derivative .operations, is recommended. Expressions for the coefficients of a low-order IIR blur filter with repeated poles are provided, as a function of scale; discrete Butterworth (IIR), and colored Savitzky-Golay (FIR), blurs are also examined. Parallel software implementations on central processing units (CPUs) and graphics processing units (GPUs), for scale-selective blob-detection in aerial surveillance imagery, are analyzed. It is shown that recursive IIR filters are significantly faster than non-recursive FIR filters when detecting large objects at coarse scales, i.e. using filters with long impulse responses; however, the margin of outperformance decreases as the degree of parallelization increases

**Keywords**: computer vision, FIR and IIR filters, image processing, object detection.

**\*E-mail:** Hugh.Kennedy@dst.defence.gov.au

## 1    Introduction

This paper is concerned with the extraction of local shape information from digital imaging sensors that are assumed to have a wide field-of-view, a fine spatial/angular resolution, and a high frame-rate; fast processing is therefore required to handle the data deluge. Maximizing the amount of data processing that is done by linear filters, with either a finite impulse response (FIR) or an infinite impulse response (IIR), is a worthwhile strategy because they only use simple shift (i.e. delay or advance), multiply and add operations. Linear filters are readily parallelized because complicated control-flows and data-transfers are not required; and their behavior is readily analyzed, characterized, and visualized, using linear-systems theory (see Appendix). Furthermore, linearity ensures that their behavior is deterministic and independent of the input data characteristics, which makes them well-suited to real-time applications. The properties of FIR and IIR filters are complementary in many respects (see Appendix); however, the image-processing literature focuses almost exclusively on FIR designs.

Discretization of a continuous prototype is a common approach in (multi-dimensional) digital-filter design because it allows a range of well-known relationships in older analog systems to be directly applied in newer digital systems[1]. However, an ideal response that is optimal in a continuous space, may not be optimal in a discrete space, although it is usually close[2,3].

Gaussian prototypes are convenient and popular in image processing because the product of the response variance in the pixel domain and the frequency domain is unity, which makes it easy to reach the desired balance between filter scale and bandwidth using the Gaussian variance ($\sigma^2$). Moreover, the product or convolution of two 1-D Gaussian responses in orthogonal dimensions with the same variance, yields a 2-D Gaussian response in the pixel/frequency domain, which has perfect rotational symmetry (isotropic), due to the summation of the squared powers of the exponential, i.e. $r^2 = x^2 + y^2$ and $r_\omega^2 = \omega_x^2 + \omega_y^2$. Indeed, this property of separable isotropy is unique to Gaussian responses[4]. However, discrete Gaussians – i.e. sampled and truncated for a non-recursive FIR filter – are only approximately isotropic in the frequency domain, with isotropy improving as dc is approached.

Recursive FIR filters that approximate discrete Gaussians have been proposed, such as sum-of-truncated-cosines[5,6] or B-splines formed from cascaded pulses[7-9]. Both approaches use the cancellation of marginally stable poles by zeros, which involve running sums realized via sliding windows, that may involve small differences of large numbers, thus they are susceptible to rounding errors. Once rounding (or other) errors accumulate, there is no inbuilt mechanism that guarantees their decay, therefore such approaches are usually avoided in online mission-critical systems built using finite-precision floating-point machines.

Recursive IIR filters that approximate discrete Gaussians have also been proposed[10-14], for example, Deriche's optimized least-squares formulations that yield closed-form expressions for filter coefficients as a function of $\sigma$, for Gaussian derivatives up to 2nd order, using IIR filters up to 4th order[10,11]. For these filters, all poles are well within the unit circle for greatly improved margins of stability and numerical robustness[10-14].

The premise of making recursive (FIR[5-9] or IIR[10-14]) approximations, of a discrete approximation, of a continuous response, is however challenged in this paper. Instead, an attempt is made to aim directly for an optimal digital filter with the desired response *ab-initio*, considering the effects of sampling (for both FIR and IIR) and truncation (for FIR only) from the outset. This then raises the question: What is "optimal" in an image-processing context? The theoretical foundations of digital filter design were established decades ago for 1-D signals, where the frequency response is the primary concern. In image processing, the impulse response is arguably more important due to the non-stationarity of the 2-D signals involved (caused by object boundaries) and a general lack of periodicity in most scenes.

Local polynomial models are a convenient representation of very low-frequency or non-periodic signals in 1-D and they lead to simple FIR (e.g. Savitzky-Golay) and IIR (e.g. Laguerre) filters[16]. Polynomial models are also used to represent 2-D signals as an alternative to sinusoidal or wavelet representations[17]. At one extreme are methods that utilize large basis sets of high-order orthogonal polynomials (e.g. Krawtchouk and Tchebichef/Chebyshev), constructed using recurrence relations, to expand an entire image frame around a single point at (or near) its center[18-21]. Smaller basis sets may be used for selected regions-of-interest, which may be non-overlapping, abutting or overlapping[22]. At the other extreme are sliding window formulations with FIR realizations, which are reached as a limiting case, for uniformly-spaced regions-of-interest of equal size that are centered on adjacent pixels[23,24]. These analysis methods are not usually considered to be filters as such and like 2-D Savitzky-Golay (SG) filters they lack the flexibility to balance scale and isotropy requirements. Furthermore, recursive methods effectively use marginally stable poles due to integrating elements. Polynomials in polar coordinates (e.g. Zernike) are ideal for isotropy[18,25,26]; however, the focus here is on Cartesian coordinates for separability and speed, due to $\mathcal{O}(2M)$ complexity instead of $\mathcal{O}(M^2)$ in the non-recursive FIR case (with a kernel length of $M$ pixels). Continuous orthogonal polynomials are also avoided to ensure discrete orthonormality[18].

The FIR and IIR filters presented in this paper are designed around 2-D polynomial models. In Section 2, the structure of this signal model is defined; then in Section 3, the use of digital differentiators to estimate the model parameters is discussed and the need for a scale-selective low-pass element is established. The use of (FIR) low-pass filters, designed using Gaussian functions, is also considered in Section 3; however, it is shown that these commonly used filters are not ideal for this purpose because they have non-vanishing moments. Therefore, in Section 4 more suitable (FIR and IIR) low-pass blur filters are proposed, that do satisfy the vanishing moment requirement, for the improved estimation of local polynomial coefficients, when cascaded with a bank of simple (FIR) high-pass differentiating filters. The properties of the combined band-pass filters are then examined in Section 5 using simulated and real images, processed using serial and parallel implementations; the main theoretical contributions, and conclusions drawn from the experiments are summarized in Section 6.

## 2 Signal model

An ideal two-dimensional (2-D) monochrome image $I$, near a spatial reference-point $(x, y)$, is represented using a bivariate Taylor-series expansion of order $D - 1$, with

$$J_{x,y}(\tilde{x}, \tilde{y}) = \sum_{d_x=0}^{D-1} \sum_{d_y=0}^{D-d_x-1} \beta_{d_x,d_y}(x,y)(\tilde{x}-x)^{d_x}(\tilde{y}-y)^{d_y} \qquad (1)$$

where

$$\beta_{d_x,d_y}(x,y) = \mathcal{D}_{d_x,d_y}(x,y)/d_x! \, d_y! \qquad (2)$$

$$\mathcal{D}_{d_x,d_y}(x,y) = \partial^{d_x+d_y} I(x,y)/\partial x^{d_x} \partial y^{d_y}. \qquad (3)$$

The $\beta_{d_x,d_y}$ coefficients (elements of the $\boldsymbol{\beta}$ matrix) of the resulting bivariate polynomial are a rich representation of local structure[27]. The index dependency ($d_x + d_y = L_d$) limits the summation to the minimum set required for rotational invariance. Knowing the partial derivatives $\mathcal{D}_{d_x,d_y}(x,y)$ up to $L_d$th-order at a given point $(x,y)$, therefore allows shape and orientation of features in the image around $(x,y)$, to be determined. These structure parameters (i.e. $\beta$ or $\mathcal{D}$) may be used in coding algorithms[17], as feature descriptors in SIFT-type algorithms[28], or as inputs to neural networks[29].

Arbitrarily-oriented derivatives $\vec{\mathcal{D}}_{\vec{d}_x,\vec{d}_y}$ at $(x,y)$ that are rotated by an angle of $\varphi$, are evaluated by substituting $\tilde{x} = \vec{x}\cos\varphi + \vec{y}\sin\varphi$ and $\tilde{y} = -\vec{x}\sin\varphi + \vec{y}\cos\varphi$ in Eqn. (1). Expanding and collecting like terms reveals that the "steered" $\vec{\beta}_{\vec{d}_x,\vec{d}_y}$ element is simply a linear combination of "raw" $\beta_{d_x,d_y}$ elements for which $d_x + d_y = \vec{d}_x + \vec{d}_y$. Thus, steered filter responses are efficiently synthesized from one set of raw filter outputs, without re-designing and re-applying custom filters for all orientations of interest. Such steering theorems are routinely used in low-level image analysis[30].

The signal model is a summation of bivariate monomials $\tilde{x}^{d_x}\tilde{y}^{d_y}$. The $\beta$ parameters are estimated using separable 2-D filters that are realized via two sequential 1-D filtering operations, e.g. applied in the $x$ then $y$ dimensions. It will be shown that under certain conditions, i.e. when moments vanish, a bank of $D$ derivative filters may be used to directly estimate the $\beta$ parameters. Our immediate objective is therefore to design a bank of $D$ filters to compute the $d$th derivative (for $0 \leq d < D$) in 1-D (e.g. $x$ or $y$). Separability reduces the number of operations required to generate the $\beta_{d_x,d_y}$ elements of the $\boldsymbol{\beta}$ matrix at each pixel using a $D \times D$ bank of filters.

## 3 Digital filters with non-vanishing moments

The fundamentals of digital differentiator design are introduced in this section; basic requirements are stated and the shortcomings of commonly used Gaussian-based methods are discussed. The Appendix should be consulted for an overview of the underlying digital systems-and-signals theory that defines the relationships between the discrete-time transfer-function (TF) $\mathcal{H}(z)$, frequency response (FR) $H(\omega)$, impulse response (IR) $h(m)$, and difference equation (DE) of a digital filter[31].

### 3.1 Differentiators

The Laplace transform of an $d$th-order differentiator in a continuous 1-D space of infinite extent is $s^d$; its FR is therefore $s^d|_{s=i\omega} = (i\omega)^d$. Corresponding derivatives at dc ($\omega = 0$) are

$$\left\{ \frac{\partial^l}{\partial \omega^l}(i\omega)^d \right\}\Big|_{\omega=0} = i^d \left\{ \frac{\partial^l}{\partial \omega^l} \omega^d \right\}\Big|_{\omega=0} = i^d d! \, \delta_{d,l} \tag{4}$$

where $\delta_{d,l}$ is the Kronecker delta

$$\delta_{d,l} = \begin{cases} 0, & \text{for } l \neq d \\ 1, & \text{for } l = d \end{cases}. \tag{5}$$

The $\rho_{d,l}^{\text{dc}}$ parameters

$$\rho_{d,l}^{\text{dc}} = \left\{ \frac{\partial^l}{\partial \omega^l} H_d(\omega) \right\}\Big|_{\omega=0} \tag{6}$$

for $0 \le l < L$, compactly characterize the response of digital differentiators to monomials in the spatial $n$-domain via Eqn. (2). The FR elsewhere, as defined by the filter zeros and poles, implicitly determines the shape of the taper or window that is applied on analysis.

Using $L = d + 1$ imposes a set of *essential* constraints on the FR of a $d$th order differentiating filter. For a bank of $D$ differentiating filters, it is *desirable* to use $L = D$ so that in the $n$-domain the $d$th filter only responds to the $d$th-order monomial and ignores all other monomials less than $D - 1$ that are considered by the other $D - 1$ filters. The principle of "vanishing moments" formalizes the link between monomials in the $n$-domain and derivatives in the $\omega$-domain[32] and it forms the foundation of the design procedure used here. Vanishing moments are usually used to derive wavelet bases for the efficient creation of a multi-scale image pyramid, which may then be traversed to find objects of unknown size; however, it is assumed here that target size is known *a-priori*; therefore, filters may be tuned to a single scale.

If $\rho_{d,l}^{\mathrm{dc}} = i^d d! \, \delta_{d,l}$, for $0 \le d < D$ and $0 \le l < D$ (in the $\omega$-domain) then

$$\frac{1}{d!} \sum_{m=-\epsilon}^{\epsilon} m^l h_d(-m) = \delta_{d,l} \text{ (in the } n\text{-domain)} \tag{7}$$

where $\epsilon = K$ and $\epsilon = \infty$ for FIR and IIR filters, respectively. Thus, polynomial coefficients may be obtained using derivative filters designed in the frequency domain instead of using (weighted) least-squares regression and orthogonal polynomials[16], or observers[33], in the pixel domain. Vanishing moments ensure that monomials and IRs are orthogonal in each dimension. Unfortunately, orthogonality and isotropy are incompatible in separable filters because Gaussian functions do not have vanishing moments, thus an appropriate balance must be found: Orthogonality ensures that 1-D low-degree polynomials in the low-frequency "passband" are handled correctly, whereas wideband isotropy ensures that all 2-D signals are handled reasonably.

In a discrete 1-D space of finite extent (that is assumed to be infinite, so that initialization transients may be ignored for the time being) we aim to reproduce this ideal response, i.e. $H_d(\omega) = (i\omega)^d$, over a narrow interval (around $\omega = 0$) with a width that is inversely proportional to scale.

In continuous and discrete domains alike, there are many challenges that must be met to realize differentiating filters with the desired response. In both cases: Frequencies beyond an upper limit are of little interest, thus the resulting filter has a band-limited response; and a taper of some kind is required to focus the response in space ($n$) with a reciprocal broadening of the response in frequency ($\omega$).

Additional complications arise in the digital case: Firstly, the sampling frequency imposes a limit on the measurable and manipulable frequencies, thus we are ignorant of what lies beyond half the sampling frequency and in between the sampled pixels. Assumptions regarding the frequency content of the signal may be used to spatially interpolate via the inverse discrete-time (or space) Fourier transform (DTFT); but in the frequency domain, we can only assume that there are no frequency components beyond $\pm\pi$, which is reasonable if the sampling period ($T_s$) is set appropriately. Secondly, the scale of analysis, which may now be finite, imposes a limit on the frequency resolution.

Simple FIR differentiators are usually designed using the standard central-difference formulae which determine the $d$th derivative of an interpolating polynomial through $2K + 1$ uniformly-spaced samples, then evaluate it at the centre of the interval where $m = 0$. The discrete-time TF of this operation is readily obtained by placing: $d$ zeros at $z = 1$ and $\Delta_d$ zeros at $z = -1$, where $\Delta_d$ is the filter parity, which is 0 and 1 for even and odd $d$, i.e. symmetric and anti-symmetric responses, respectively.

$$\mathcal{H}_d^{\text{hpf}}(z) = \frac{\mathcal{B}(z)}{\mathcal{A}(z)} = \frac{(z-1)^d(z+1)^{\Delta_d}}{(\Delta_d+1)z^{K^{\text{hpf}}}}. \tag{8}$$

The IR, i.e. the standard central-difference formulae of arbitrary order (see TABLE I. ), is then determined using $h_d^{\text{hpf}}(m) = \mathcal{Z}^{-1}\{\mathcal{H}_d^{\text{hpf}}(z)\}$. As all poles are at $z = 0$, the filter coefficients $b_m^{\text{hpf}}$ and $a_m^{\text{hpf}}$ for $K^{\text{hpf}} = (d + \Delta_d)/2$, are simply set equal to the coefficients of the $\mathcal{B}(z)$ and $\mathcal{A}(z)$ polynomials, or factored into the three-part DE using Eqn. (A17). The $\rho_{d,l}^{\text{dc}}$ parameters (see TABLE II. ) for this simple differentiator indicate that the interpolating polynomials are orthogonal for $D = 3$ (due to symmetry); however, this is not true in the general case. For example, in the $D = 5$ case, non-vanishing moments give rise to ambiguities because the $d = 1$ filter responds to both the $m^1$ and $m^3$ monomials; similarly, the $d = 2$ filter responds to the $m^2$ and $m^4$ monomials.

This wide-band $d$th-order interpolating differentiator $\mathcal{H}_d^{\text{hpf}}(z)$, acts as a high-pass filter (hpf), for $d > 0$ and for $\Delta_d = 0$, in particular (see Fig. 1, left); therefore, it is cascaded with a low-pass filter (lpf) with TF $\mathcal{H}_0^{\text{lpf}}(z)$, to attenuate out-of-band signals (e.g. clutter at fine scales) and noise, to yield a band-pass filter (bpf) with TF $\mathcal{H}_d(z) = \mathcal{H}_0^{\text{lpf}}(z)\mathcal{H}_d^{\text{hpf}}(z)$ (see Fig. 1, right and Fig. 2). Low-pass FIR filters derived from Gaussian functions are commonly used for this purpose. As long IRs are required for coarse-scale analysis, using this two-stage configuration with a common Gaussian blur (a high-order FIR lpf) and a parallel bank of interpolating differentiators (low-order FIR hpfs), significantly reduces the number of arithmetic operations. For best results, the Gaussian should be discretized using a large value of $K^{\text{lpf}}$, for high-fidelity reproduction of the tails, because the high gain of $\mathcal{H}_d^{\text{hpf}}(z)$ amplifies any imperfections in $\mathcal{H}_0^{\text{lpf}}(z)$, e.g. high sidelobes of $H_0^{\text{lpf}}(\omega)$ or sharp discontinuities in $h_0^{\text{lpf}}(m)$.

TABLE I.    IMPULSE RESPONSE $h_d^{\text{hpf}}(m)$ OF INTERPOLATING DERIVATIVE FILTERS.

| | $m = -2$ | $m = -1$ | $m = 0$ | $m = 1$ | $m = 2$ |
|---|---|---|---|---|---|
| **d = 0** | 0.0 | 0.0 | 1.0 | 0.0 | 0.0 |
| **d = 1** | 0.0 | 0.5 | 0.0 | -0.5 | 0.0 |
| **d = 2** | 0.0 | 1.0 | -2.0 | 1.0 | 0.0 |
| **d = 3** | 0.5 | -1.0 | 0.0 | 1.0 | -0.5 |
| **d = 4** | 1.0 | -4.0 | 6.0 | -4.0 | 1.0 |

TABLE II.    VALUES OF $\rho_{d,l}^{\mathrm{dc}}$ FOR FREQUENCY RESPONSE $H_d^{\mathrm{hpf}}(\omega)$ OF INTERPOLATING DERIVATIVE FILTERS.

| | **l = 0** | **l = 1** | **l = 2** | **l = 3** | **l = 4** |
|---|---|---|---|---|---|
| **d = 0** | 1 | 0 | 0 | 0 | 0 |
| **d = 1** | 0 | $i$ | 0 | $-i$ | 0 |
| **d = 2** | 0 | 0 | -2 | 0 | 2 |
| **d = 3** | 0 | 0 | 0 | $-6i$ | 0 |
| **d = 4** | 0 | 0 | 0 | 0 | 24 |

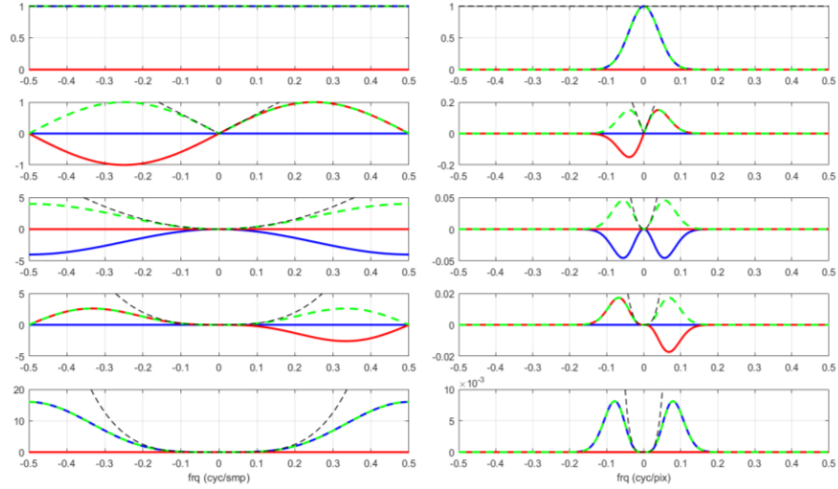

Fig. 1.  1-D frequency response of $H_d^{\mathrm{hpf}}(\omega)$ (left) and $H_d^{\mathrm{bpf}}(\omega)$ (right); for $d = 0$ (top) to $d = 4$ (bottom) over $f = \pm 0.5$ (cycles per pixel). Real (solid blue) and imaginary (solid red) parts shown with magnitude (dashed green) and magnitude of continuous fullband differentiator (dashed black) on a linear scale. Gaussian lpf configured with $\sigma = 4$ and $K = 5\sigma$.

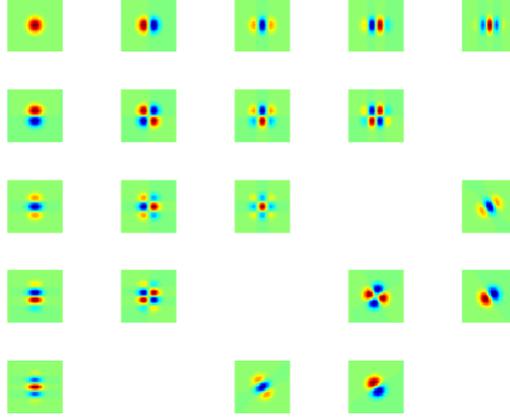

Fig. 2. In upper-left triangle: 2-D impulse response $h^{\text{bpf}}_{d_x,d_y}(m)$ of a Gaussian bpf for $d_x = 0 \dots 4$ (top to bottom) and $d_y = 0 \dots 4$ (left to right). Positive parts in red; negative in blue. Gaussian lpf configured with $\sigma = 4$ and $K = 5\sigma$. In lower-right triangle: $h^{\text{bpf}}$ rotated/steered by/to 30°; $h^{\text{bpf}}_{1,0}$ and $h^{\text{bpf}}_{0,1}$ along first diagonal; $h^{\text{bpf}}_{2,0}$, $h^{\text{bpf}}_{1,1}$ and $h^{\text{bpf}}_{0,2}$ along second diagonal.

### 3.1 Gaussian blurs and differentiators

Differentiators with an integrated low-pass element may also be realized in a one-stage configuration as an alternative to the separated two-stage configuration recommended in the previous sub-section by taking the $d$th derivative of a (continuous) Gaussian function $g(x)$ with a standard deviation of $\sigma$ pixels, then sampling the result (a Hermite function) at the required rate over the desired extent[17,27,34-37]. A Gaussian blur is obtained in the limiting case, using $d = 0$; for the general case ($d \geq 0$), in the $x$ dimension,

$$h_{d_x}(m_x) = c_g g_{d_x}(x)\big|_{x=m_x T_s} \text{ where} \tag{9a}$$

$$g_{d_x}(x) = \frac{\partial^{d_x}}{\partial x^{d_x}} g_0(x) \text{ and} \tag{9b}$$

$$g_0(x) = \frac{1}{\sigma\sqrt{2\pi}} e^{-x^2/2\sigma^2}. \tag{9c}$$

The FR of the ideal analogue prototype $g_0(x)$, obtained via the Fourier transform, is also Gaussian

$$G_0(\omega_x) = e^{-\sigma^2 \omega_x^2/2}. \tag{10}$$

On discretization, re-normalization of the differentiator response for unity gain at dc, such that $H_0(\omega_x)|_{\omega_x=0} = 1$, is also required. This is conveniently done in the $n$-domain using

$$\frac{1}{c_g} = H_0(\omega_x)|_{\omega_x=0} = \sum_{m_x=-K}^{K} h_0(m_x). \tag{11}$$

After discretization, the "tails" of the (finite) IR $h_{d_x}(m_x)$ and its FR $H_{d_x}(\omega_x)$ are truncated due to a finite $K$ and a non-zero $T_s$, respectively. As multiplication in the spatial domain is convolution in the frequency domain, its FR is also convolved with the Dirichlet kernel (sometimes known as the periodic sinc function), which is the DTFT of the rectangular window: $w(m_x) = 1$ for $m_x = -K \ldots K$ and $w(m_x) = 0$, elsewhere. This convolution results in a broadening of the mainlobe centered on $\omega_x = 0$ for $d_x = 0$, and the introduction of sidelobes. The width of the broadening and the height of the sidelobes increase (for degraded frequency selectivity) as the width of the window decreases (for improved spatial selectivity and computational efficiency).

The $l$th derivative of the FR for the $d$th filter at dc, i.e. $\rho_{d,l}^{\mathrm{dc}}$, are tabulated in TABLE III. , for some Gaussian filters. Responses of the continuous prototype $G_d(\omega)$, as a function of the $\sigma$ parameter, are provided; the realized responses of the corresponding discrete filters, using $\sigma = 1$ pixel (pix), are shown using floating-point values; cascaded two-stage responses are shown in italics. For the two-stage realization, with $K^{\mathrm{lpf}} = 5$, the discrete values are equal to the continuous values (to more than 4 decimal places) for $l \leq d$, where they matter most. For the one-stage realization, with $K = 5$, the discrete values are approximately equal to the continuous values – the error increases with both $l$ and $d$; however, this may be offset by increasing $K$ at the expense of computational complexity. Errors also increase with decreasing $\sigma$ and this can only be offset by decreasing $T_s$. Symmetry ensures that derivatives vanish exactly for $l$ and $d$ of different parity, for $G_d(\omega)$ and $H_d(\omega)$ alike. Note also that even in the continuous case, a $d$th-order differentiator has a non-vanishing response for $l > d$, when $l$ and $d$ have the same parity. Consequently, responses are ambiguous for polynomial components of like symmetry if these filters are used directly without orthogonalization[34,35]; although, in Ref. 17 it is claimed that this "redundancy" need not be of concern.

TABLE III.  Values of $\rho_{d,l}^{dc}$ for Gaussian derivative filters. Analog prototype as a function of $\sigma$; digital one-stage and two-stage realizations, in roman and italics floating-point respectively, using $\sigma = 1$.

|  | $l = 0$ | $l = 1$ | $l = 2$ | $l = 3$ | $l = 4$ |
|---|---|---|---|---|---|
| $d = 0$ | **1** | 0 | $-\sigma^2$ | 0 | $3\sigma^4$ |
|  | 1.0000 + 0.0000i | 0.0000 + 0.0000i | -1.0000 + 0.0000i | 0.0000 + 0.0000i | 3.0000 + 0.0000i |
|  | *1.0000 + 0.0000i* | *0.0000 + 0.0000i* | *-1.0000 + 0.0000i* | *0.0000 + 0.0000i* | *3.0000 + 0.0000i* |
| $d = 1$ | 0 | $i$ | 0 | $-3i\sigma^2$ | 0 |
|  | 0.0000 + 0.0000i | 0.0000 + 1.0000i | 0.0000 + 0.0000i | 0.0000 - 3.0000i | 0.0000 + 0.0000i |
|  | *0.0000 + 0.0000i* | *0.0000 + 1.0000i* | *0.0000 + 0.0000i* | *0.0000 - 4.0000i* | *0.0000 + 0.0000i* |
| $d = 2$ | 0 | 0 | **−2** | 0 | $12\sigma^2$ |
|  | -0.0000 + 0.0000i | 0.0000 + 0.0000i | -2.0000 + 0.0000i | 0.0000 + 0.0000i | 11.9992 + 0.0000i |
|  | *-0.0000 + 0.0000i* | *0.0000 + 0.0000i* | *-2.0000 + 0.0000i* | *0.0000 + 0.0000i* | *14.0000 + 0.0000i* |
| $d = 3$ | 0 | 0 | 0 | $-6i$ | 0 |
|  | 0.0000 + 0.0000i | 0.0000 + 0.0000i | 0.0000 + 0.0000i | 0.0000 - 5.9992i | 0.0000 + 0.0000i |
|  | *0.0000 + 0.0000i* | *0.0000 + 0.0000i* | *0.0000 + 0.0000i* | *0.0000 - 6.0000i* | *0.0000 + 0.0000i* |
| $d = 4$ | 0 | 0 | 0 | 0 | **24** |
|  | 0.0000 + 0.0000i | 0.0000 + 0.0000i | 0.0000 + 0.0000i | 0.0000 + 0.0000i | 23.9896 + 0.0000i |
|  | *0.0000 + 0.0000i* | *0.0000 + 0.0000i* | *0.0000 + 0.0000i* | *0.0000 + 0.0000i* | *24.0000 + 0.0000i* |

## 4 Digital filters with vanishing moments

Frequency-domain procedures for the design of differentiating filters with vanishing moments (for monomial selectivity) and adjustable bandwidth (for scale selectivity) are described in this section. A two-stage configuration with serial lpf-blur and hpf-differentiator stages is used for a bpf bank. It is relatively straightforward to design a bank of $D$ low-order FIR filters that satisfy the necessary $D$ derivative constraints at dc, for vanishing moments in the pixel domain. Furthermore, if the frequency-domain derivatives of the lpf blur are flat up to order $D - 1$ at dc, then the lpf and the hpf may be designed independently without the introducing non-vanishing moments in the combined bpf. These low-pass and high-pass filters (respectively) assume the roles of the interpolator $p[\cdot]$ and the differentiator $d[\cdot]$ in Ref. 38. The shape of the taper (or interpolator) in the $n$- and $\omega$-domains is simply parameterized for the filters proposed here without the numerical optimizations used in Ref. 38. The low-pass operation also resembles the role played by scaling-function in wavelet decompositions[32].

In this section, a general procedure for the design of a FIR or IIR filter that satisfies derivative constraints at dc and Nyquist (i.e $\omega = 0$ and $\omega = \pi$) is presented. Orthogonality and monomial selectivity in the pixel domain is the primary objective; response isotropy is an important but secondary consideration. The procedure is then adapted and customized to design: FIR differentiators and IIR repeated-pole blurs. Procedures for the design of FIR blurs using colored Savitzky-Golay smoothing and IIR blurs designed using an analog Butterworth prototype are also presented.

The objective is to design a bank of $D$ derivative filters, where the $d$th filter (for $0 \le d < D$), with FR $H_d(\omega)$, has the desired derivatives at $\omega = 0$ and $\omega = \pi$, i.e. $\rho_{d,l}^{\text{dc}} = i^d d! \, \delta_{d,l}$ and $\rho_{d,l}^{\text{pi}} = 0$, for $0 \le l < L^{\text{dc}}$ and $0 \le l < L^{\text{pi}}$, respectively, with $L^{\text{dc}} = D$. As an example, for $D = 5$, we would like the $l = d$ derivatives of the FR at dc to equal the values in bold in TABLE III. and all other $l \ne d$ derivatives at dc to be zero. The proposed method is an extension of Ref. 39; the derivation below is simplified by the absence of the delay parameter $q$, which is assumed here to be zero; but complicated by the need to consider non-causal terms (see Appendix).

The $d$th filter response is expressed as a linear combination of $N_K$ basis functions, such that

$$h_d(m) = \sum_{k=0}^{N_K - 1} c_k f_k(m) \tag{12a}$$

$$\mathcal{H}_d(z) = \sum_{k=0}^{N_K - 1} c_k \mathcal{F}_k(z) \text{ and} \tag{12b}$$

$$H_d(\omega) = \sum_{k=0}^{N_K - 1} c_k F_k(\omega). \tag{12c}$$

The coefficients $c_k$, are found by solving the system of linear equations $\boldsymbol{\rho} = \boldsymbol{F} \boldsymbol{c}$ using $\boldsymbol{c} = \boldsymbol{F}^{-1} \boldsymbol{\rho}$, where $\boldsymbol{c}$ and $\boldsymbol{\rho}$ are column vectors of length $N_K$ containing the unknown coefficients and the $\omega$-domain derivative constraints, respectively. In the general case, the coefficients of the non-realizable filter are found by substituting $\boldsymbol{c}$ into Eqn. (12b) and expanding for $\mathcal{B}(z)$ and $\mathcal{A}(z)$; the DE coefficients of the realizable filter are then found using Eqn. (A17).

Symmetry automatically ensures that all derivatives of the FR vanish exactly when the parity of $l$ does not match the parity of $d$. These derivatives do not need to be considered explicitly, thus the number of constraints and unknowns are reduced accordingly. Using odd $D$ is recommended so that, excluding $d = 0$, there are an equal number of $d$ values with zero and unity parity; thus $D = 2L_D + 1$, where $L_D$ is the number of parity pairs. This simplifies the filter design logic and facilitates shape analysis. Let $\bar{L}^{\text{dc}}$ and $\bar{L}^{\text{pi}}$ be the number of $\omega = 0$ and $\omega = \pi$ constraints after the consideration of symmetry; thus the length of $\boldsymbol{\rho}$ and $\boldsymbol{c}$, i.e. $\bar{L}$ and $N_K$, is equal to $\bar{L}^{\text{dc}} + \bar{L}^{\text{pi}}$. For the $d$th filter, $\bar{L}^{\text{dc}} = L_D - \Delta_d + 1$, which is equal to the number of $l$ values that have the required parity on the interval $[0, D - 1]$; $\bar{L}^{\text{pi}}$ is independent of $d$. For parity pairs of the same order, the anti-symmetric filter (i.e. odd $d$ with $\Delta_d = 1$) has one less degree of freedom available because the response is constrained to be zero at $m = 0$ and $\omega = 0$, but it also has one less derivative constraint that it must satisfy. Parity pairs, for filters with a FIR, have the same length $M = 2K + 1$ (thus order), where $K = L_D + \bar{L}^{\text{pi}}$. Note that $\boldsymbol{\rho}$, $\boldsymbol{c}$ and $\boldsymbol{F}$ are all indexed from zero; furthermore, the rows of $\boldsymbol{\rho}$ and $\boldsymbol{F}$ are indexed using $\bar{l}$, with $l = 2\bar{l} + \Delta_d$, because $l$-values with parity that does not match $d$ are omitted.

Two vectors are stacked to form $\boldsymbol{\rho}$

$$\boldsymbol{\rho} = \begin{bmatrix} \boldsymbol{\rho}^{\text{dc}} \\ \boldsymbol{\rho}^{\text{pi}} \end{bmatrix} \text{ where} \tag{13}$$

$\boldsymbol{\rho}^{\mathrm{dc}}$ is a column vector of length $\bar{L}^{\mathrm{dc}}$ with the $\bar{l}$th element equal to $\rho_{d,l}^{\mathrm{dc}}$ and $\boldsymbol{\rho}^{\mathrm{pi}}$ is a column vector of length $\bar{L}^{\mathrm{pi}}$ with the $\bar{l}$th element equal to $\rho_{d,l}^{\mathrm{pi}}$, thus $\boldsymbol{\rho}$ contains all zeros, except for the $\bar{l} = (d - \Delta_d)/2$ element, which is equal to $i^d d!$.

Similarly, $\boldsymbol{F}$ is formed by stacking two matrices

$$\boldsymbol{F} = \begin{bmatrix} \boldsymbol{F}^{\mathrm{dc}} \\ \boldsymbol{F}^{\mathrm{pi}} \end{bmatrix} \text{ where} \tag{14}$$

$\boldsymbol{F}^{\mathrm{dc}}$ and $\boldsymbol{F}^{\mathrm{pi}}$ are $\bar{L}^{\mathrm{dc}} \times N_K$ and $\bar{L}^{\mathrm{pi}} \times N_K$ matrices, with elements in the $\bar{l}$th row and the $k$th column

$$F_{\bar{l},k}^{\mathrm{dc}} = \left\{ \frac{\partial^l}{\partial \omega^l} F_k(\omega) \right\} \Big|_{\omega=0} \text{ and} \tag{15a}$$

$$F_{\bar{l},k}^{\mathrm{pi}} = \left\{ \frac{\partial^l}{\partial \omega^l} F_k(\omega) \right\} \Big|_{\omega=\pi} \text{ using} \tag{15b}$$

$l = 2\bar{l} + \Delta_d$, in accordance with symmetry.

The form of the basis functions is determined by the filter configuration, i.e. one-stage or two-stage, and the IR duration, i.e. finite (realized non-recursively) or infinite (realized recursively). They are selected so that their symmetry matches the symmetry of the filter. In practice the number of constraints used in the design is limited by the conditioning of $\boldsymbol{F}$; which may be difficult to accurately invert for coarse scales and large $\bar{L}$.

*4.1 FIR derivative filters*

The TF in Eqn. (8) has vanishing moments for $0 \le l < d$, the desired response for $l = d$ but the moments for $d < l < D$ and $\Delta_l = \Delta_d$, do not vanish. For FIR filters, this deficiency is rectified using the procedure described above with a basis set consisting of pure delays, i.e.

$$f_k(m) = \mu_k \left\{ \delta_{-k-\Delta_d,m} + (-1)^{\Delta_d} \delta_{k+\Delta_d,m} \right\} \tag{16a}$$

$$\mathcal{F}_k(z) = \mathcal{Z}\{f_k(m)\} = \mu_k \left\{ z^{-(k+\Delta_d)} + (-1)^{\Delta_d} z^{(k+\Delta_d)} \right\} \text{ and} \tag{16b}$$

$$F_k(\omega) = \mathcal{F}_k(z)|_{z=e^{i\omega}} = \mu_k\{e^{-i(k+\Delta_d)\omega} + (-1)^{\Delta_d}e^{i(k+\Delta_d)\omega}\} \text{ where} \tag{16c}$$

$$\mu_k = 2^{-(1-\Delta_d)\delta_{k,0}}. \tag{16d}$$

The normalizing factor $\mu_k = 1/2$ when $k = 0$ and $\Delta_d = 0$, i.e. when the both unit impulses in Eqn. (16a) coincide and sum at $m = 0$; otherwise $\mu_k = 1$.

In this FIR case, where all poles are at the origin of the $z$-plane, once $\boldsymbol{c}$ has been determined, it is simpler to extract the DE coefficients of the filter using

$$b_k^+ = c_k, \; b_k^- = c_k \text{ and } b_0^\circ = c_0, \text{ for } 0 < k < N_K, \text{ when } \Delta_d = 0 \text{ and} \tag{17a}$$

$$b_{k+1}^+ = -c_k, \; b_{k+1}^- = c_k \text{ and } b_0^\circ = 0, \text{ for } 0 \le k < N_K, \text{ when } \Delta_d = 1 \tag{17b}$$

instead of using Eqn. (A17). The DE of the FIR filter has $M^{\text{hpf}} = 2K^{\text{hpf}} + 1$ coefficients where $K^{\text{hpf}} = N_K + \Delta_d - 1$, i.e. $K^{\text{hpf}} = \bar{L}^{\text{dc}} + \bar{L}^{\text{pi}} + \Delta_d - 1$. When this hpf $\mathcal{H}_d^{\text{hpf}}(z)$, is cascaded with an lpf $\mathcal{H}_0^{\text{lpf}}(z)$, designed using large $\sigma$ and $K^{\text{lpf}}$, Nyquist constraints are unnecessary because high-frequency gain is already negligible, thus $K^{\text{hpf}} = \bar{L}^{\text{dc}} + \Delta_d - 1$.

### 4.2 FIR colored Savitzky-Golay blur filters

FIR Savitzky-Golay (SG) filters operate by least-squares fitting a polynomial of degree $D - 1$ to a finite data sequence[16]. In the $n$-domain, the fit is done in a way that minimizes the sum of squared residuals, with a uniform weighting function so that all pixels contribute equally. The $d$th derivative of the fitted polynomial is then evaluated at the center of the analysis window (where $m = 0$) in the linear-phase case. Unlike the interpolating differentiator in Section 3, $K \gg D$ is usually used for low-pass or smoothing effect. In the $\omega$-domain, this is equivalent to the minimization of the white-noise gain (WNG) of the filter's frequency response i.e.

$$\text{WNG} = \frac{1}{2\pi}\int_{-\pi}^{\pi}\left|H_0^{\text{lpf}}(\omega)\right|^2 d\omega = \frac{1}{2\pi}\left\|H_0^{\text{lpf}}(\omega)\right\|_2^2 \tag{18}$$

where $\|\cdot\|_2$ is the $\mathcal{L}_2$-norm; or, as a consequence of Parseval's theorem

$$\text{WNG} = \sum_{m=-K}^{K} \left| h_0^{\text{lpf}}(m) \right|^2 = \left\| h_0^{\text{lpf}}(m) \right\|_2^2 \tag{19}$$

subject to dc derivative constraints $\rho_{d,l}^{\text{dc}} = i^d d! \, \delta_{d,l}$, for $0 < l < D$. This is accomplished via Lagrange multipliers using

$$\begin{bmatrix} \boldsymbol{c}_{N_K \times 1} \\ \boldsymbol{\lambda}_{\bar{L}^{\text{dc}} \times 1} \end{bmatrix} = \begin{bmatrix} \boldsymbol{S}_{N_K \times N_K} & \left( \boldsymbol{F}_{\bar{L}^{\text{dc}} \times N_K}^{\text{dc}} \right)^T \\ \boldsymbol{F}_{\bar{L}^{\text{dc}} \times N_K}^{\text{dc}} & \boldsymbol{0}_{\bar{L}^{\text{dc}} \times \bar{L}^{\text{dc}}} \end{bmatrix}^{-1} \begin{bmatrix} \boldsymbol{0}_{N_K \times 1} \\ \boldsymbol{\rho}_{\bar{L}^{\text{dc}} \times 1}^{\text{dc}} \end{bmatrix} \tag{20}$$

where $\boldsymbol{S}$, $\boldsymbol{F}$ and $\boldsymbol{\rho}$ are populated using a basis set of $N_K$ unit impulses. In the uniformly-weighted/white-noise case, due to the orthonormality of $f_k(m)$ and $F_k(\omega)$, we have $\boldsymbol{S} = \boldsymbol{I}$. SG filters are a convenient starting point but without modification they do not provide a mechanism for the selection of scale. For a given $D$, the WNG decreases as $K$ increases, due to a lowering of the sidelobes and a narrowing of the transition band. The bandwidth, however, remains approximately constant. This deficiency is addressed here by minimizing the colored-noise gain (CNG), where

$$\text{CNG} = \frac{1}{\pi} \int_{\omega_c}^{\pi} \left| H_0^{\text{lpf}}(\omega) \right|^2 d\omega \tag{21}$$

subject to dc derivative constraints $\rho_{d,l}^{\text{dc}} = i^d d! \, \delta_{d,l}$, for $0 \leq l < D$, and where $d = 0$, for the (symmetric) blur filters considered here. Minimizing the CNG, instead of the WNG, obfuscates the interpretation of this filter as a weighted least-squares fitting operation as the error weighting function and the form of the orthogonal polynomials are unknown.

The elements of $\boldsymbol{S}$ are now evaluated using $\boldsymbol{S} = \boldsymbol{\mathcal{T}} \bar{\boldsymbol{S}} \boldsymbol{\mathcal{T}}^T$ where the elements of $\bar{\boldsymbol{S}}$ (a $M$ by $M$ matrix, indexed from $0$ to $M-1$, with $M = 2K+1$ and $K = N_K - 1$) are definite integrals of pairwise products of complex sinusoids ($e^{im_1\omega} \times e^{-im_2\omega}$) in the $\omega$-domain (which are real due to symmetry) and where the $\boldsymbol{\mathcal{T}}$ operator (a $K+1$ by $M$ matrix) transforms the basis set of complex sinusoids into the basis set of real (symmetric) cosine functions used in $\boldsymbol{F}$ and $\boldsymbol{\rho}$, i.e. for $m_1 = -K \ldots K$ and $m_2 = -K \ldots K$:

$$\bar{S}_{K+m_1, K+m_2} = \frac{1}{2\pi} \left\{ \int_{-\pi}^{-\omega_c} e^{i(m_1-m_2)\omega} d\omega + \int_{\omega_c}^{\pi} e^{i(m_1-m_2)\omega} d\omega \right\} \tag{22a}$$

$$= \frac{1}{\pi} \times \begin{cases} \frac{\sin\{\pi(m_1-m_2)\}}{m_1-m_2} - \frac{\sin\{\omega_c(m_1-m_2)\}}{m_1-m_2} & , m_1 \neq m_1 \\ \pi - \omega_c & , m_1 = m_1 \end{cases} \quad \text{and} \tag{22b}$$

$$\mathcal{T}_{k, K+m_1} = \delta_{k, |m_1|}. \tag{22c}$$

This formulation yields a blur filter with a magnitude response that is somewhat Gaussian, with flattened tails for $\omega_c \leq |\omega| \leq \pi$ (due to CNG minimization), a flat response near $\omega = 0$ (due to constraint satisfaction), and a smooth roll-off from 0 to $\pm\omega_c$. The bandwidth parameter may now be used to determine the scale of analysis, assuming $N_K$ is sufficiently large for a reasonably low CNG. Using $K^{\mathrm{lpf}} = \lceil 5\sigma \rceil$ and $\omega_c = 3/\sigma$, i.e. 5- and 3-sigma points in the pixel and frequency domains (respectively), is sufficient for $L_D = 1$, where $\lceil \cdot \rceil$ rounds up to the nearest integer; $K^{\mathrm{lpf}} = \lceil 7\sigma \rceil$ and $\omega_c = 3/\sigma$ is recommended for $L_D = 2$. The $\sigma$ parameter is the standard deviation of the "equivalent" Gaussian in the pixel domain, to which the colored SG filter is matched.

### 4.3 IIR repeated-pole blur filters

For these low-pass blurs, the basis functions have IRs $f_k(m)$, that are computed recursively. Consequently, all poles of the TF $\mathcal{H}_0^{\mathrm{lpf}}(z)$ are shifted right along the real axis in the $z$-plane from the origin (a $K$th-order delay) and towards the unit circle (a $K$th-order integrator). When $d$ is even ($\Delta_d = 0$), the unit impulse (at $m = 0$) is also included; when $d$ is odd ($\Delta_d = 1$), the unit impulse is excluded because all basis functions are required to be zero at the IR midpoint for anti-symmetry. The corresponding TFs are derived by summing the forward part of the IR (which includes $m = 0$) and a reversed version of the forward part. The value of the IR at $m = 0$ is also subtracted so that the response there is only considered once. The forward part of the recursive basis functions has repeated poles on the real axis at $p = e^{-1/\sigma}$ with $0 < p < 1$ and a multiplicity of $k + 1$, where $\sigma$ is the mean of the one-sided exponential. The $\sigma$ parameter is used to determine the scale of analysis in the same way that $\sigma$ is used in Gaussian filters. The IR of the $k$th basis function is defined as follows:

For $k = L_D + \bar{L}^{\text{pi}}$ (i.e. when $\Delta_d = 0$ and $k = N_K - 1$), $f_k(m) = \delta_{0,m}$ (the unit impulse). Otherwise

$$f_k(m) = \begin{cases} f_k^+(m) & , m \geq 0 \\ f_k^-(m) & , m < 0 \end{cases} \text{ where} \tag{23a}$$

$$f_k^+(m) = m^k p^m \text{ and } f_k^-(m) = (-1)^{\Delta_d}(-m)^k p^{-m} . \tag{23b}$$

When $\Delta_d = 1$, anti-symmetry requires $f_k(0) = 0$ for $0 \leq k < N_K$. When $\Delta_d = 0$, no such constraint is imposed and $f_k(0) = 1$ for $k = 0$ and $k = N_K - 1$; $f_k(0) = 0$ for $0 < k < N_K - 1$.

The TF of the $k$th basis function is

$$\mathcal{F}_k(z) = \mathcal{F}_k^+(z) + (-1)^{\Delta_d}\mathcal{F}_k^+(z)|_{z=z^{-1}} - (1 - \Delta_d)f_k^+(m)|_{m=0} \tag{24}$$

where $\mathcal{F}_k^+(z)$ is computed using the one-sided $\mathcal{Z}$ transform (in the forward direction) of $f_k^+(m)$. The FR of the $k$th basis function is $F_k(\omega) = \mathcal{F}_k(z)|_{z=e^{i\omega}}$. The rows of $\boldsymbol{F}$ are built via the repeated application of the quotient rule for rational functions in $e^{i\omega}$. The order of the resulting rational function quickly grows large. When it is evaluated at dc, the result is very sensitive to the positions of its poles and zeros which may be tightly clustered around $z = 1$. Thus the order of the filter designed by this method is limited by arithmetic precision. Once the $\boldsymbol{F}$ matrix has been populated and the $c_k$ coefficients determined, $\mathcal{H}_0^{\text{lpf}}(z)$ is constructed using Eqn. (12b) and rearranged so that it is in the standard $\mathcal{B}(z)/\mathcal{A}(z)$ form. The order $(2K^{\text{lpf}})$ of $\mathcal{H}_0^{\text{lpf}}(z)$ is equal to $2(L_D + \bar{L}^{\text{pi}})$. The decomposition outlined in Eqn. (A17) is then used to extract the coefficients of the forward, backward and central parts of the DE. The forward and backward filters have $K^{\text{lpf}}$ non-zero $b_m$ coefficients and a non-zero central coefficient $b_0^\circ$.

### 4.4 IIR Butterworth blur filters

This low-pass blur filter is derived from a continuous $L_B$th-order Butterworth prototype

$$\mathcal{H}_0^{\text{lpf}}(s) = \left\{ 1 + \left( \frac{-s^2}{\omega_c^2} \right)^{\frac{L_B}{2}} \right\}^{-1} \tag{25}$$

where $L_B = 2(L_D + 1)$ for a "flatness" order of $L_B - 1$ at dc, i.e. $\rho_{0,l}^{\text{dc}} = \delta_{0,l}$ for $0 \leq l < L_B$. The blur is tuned to a scale by selecting the cutoff frequency $(\omega_c)$ so that the half-gain point of $H_0^{\text{lpf}}(\omega) = \mathcal{H}_0^{\text{lpf}}(s)\big|_{s=i\omega}$ is matched to the half-gain point of an "equivalent" Gaussian $G(\omega)$ with a $\omega$-domain variance of $1/\sigma^2$, using $\omega_c = \sqrt{2\log(2)}/\sigma$. The prototype is discretized via the bilinear transformation, $\mathcal{H}_0^{\text{lpf}}(z) = \mathcal{H}_0^{\text{lpf}}(s)\big|_{s=2\frac{z-1}{z+1}}$. Pre-warping is optional but not essential because exactly matched magnitudes at $\omega_c$ in the $s$- and $z$-domains is not critical – an approximate bandwidth for scale selectivity is sufficient. The bilinear transform preserves dc flatness and places $L_B$ zeros at $\omega = \pi$, for a flatness order of $L_B - 1$ at Nyquist as a bonus, i.e. $\rho_{0,l}^{\text{pi}} = 0$, for $0 \leq l < L_B$. DE coefficients of the realizable filter (with $K^{\text{lpf}} = L_D + 1$) are extracted using Eqn. (A17).

## 5  Application and analysis

Using $D = 3$ is sufficient for basic shape analysis in some functions, such as blob detection, as a pre-screening stage, in medical imaging and aerial surveillance applications. The resulting bivariate quadratic model for each pixel allows local maxima to be identified and basic shape parameters to be estimated via the Hessian matrix [2,7,40,41]. The Hessian matrix at $(x, y)$ is defined as follows:

$$\mathbb{H}(x, y) = \begin{bmatrix} \mathcal{D}_{2,0}(x, y) & \mathcal{D}_{1,1}(x, y) \\ \mathcal{D}_{1,1}(x, y) & \mathcal{D}_{0,2}(x, y) \end{bmatrix} \tag{26}$$

where $\mathcal{D}_{d_x, d_y}(x, y)$ are the local partial derivatives of image intensity as defined in Eqns. (1)-(3), which are computed by applying separable 1-D filters with TFs $\mathcal{H}_{d_x}^{\mathrm{bpf}}(z)$ and $\mathcal{H}_{d_y}^{\mathrm{bpf}}(z)$ in the $x$ then $y$ directions, respectively. The fitted surface has a local maximum or minimum when the determinant of Hessian matrix is greater than zero, i.e. $|\mathbb{H}| > 0$, where $|\mathbb{H}| = \mathcal{D}_{2,0}\mathcal{D}_{0,2} - \mathcal{D}_{1,1}\mathcal{D}_{1,1}$. If a bright blob is assumed to be a bivariate Gaussian centered at $(x, y)$, corresponding to a probability density function (pdf), then $-\mathbb{H}$ is its Fisher information matrix $\mathbb{F}$ and $\mathbb{F}^{-1}$ is its covariance matrix $\mathbb{P}$. This analogy provides a geometric of interpretation of a (positive semi-definite) Hessian matrix with a positive determinant. The relative length and orientation of the blob's principal axes may then be determined from an eigen-decomposition of $\mathbb{P}$. Applying a threshold to the determinant favors the detection of sharp local maxima/minima on bright/dark blobs. Using the normalized determinant of the Hessian matrix, i.e. $\sigma^4|\mathbb{H}|$, allows a single threshold (Threshold #1) to be used for all scales[41]. Invoking the pdf interpretation of the blob once again, this quantity is the squared ratio of the area of two ellipses that enclose equal probability masses; where the numerator ellipse (a circle) is formed from an isotropic pdf with a covariance matrix equal to the scale $\sigma^2 \boldsymbol{I}_{2\times 2}$, and the denominator ellipse is formed from a pdf with a covariance matrix equal to $\mathbb{P}$.

There is however no guarantee that the local maximum/minimum is at or even close to $(x, y)$. The distance to the local maximum/minimum is $\sqrt{(\Delta x)^2 + (\Delta y)^2}$ where the displacements $\Delta x$ & $\Delta y$ are computed from the locally-fitted bi-quadratic surface using

$$\Delta x = \left( \mathcal{D}_{0,1} \mathcal{D}_{1,1} - \mathcal{D}_{0,2} \mathcal{D}_{1,0} \right) / |\mathbb{H}| \text{ and} \tag{27a}$$

$$\Delta y = \left( \mathcal{D}_{1,0} \mathcal{D}_{1,1} - \mathcal{D}_{2,0} \mathcal{D}_{0,1} \right) / |\mathbb{H}|. \tag{27b}$$

Placing a limit (Threshold #2) on the distance also favors sharp local maxima/minima and is intended to suppress spurious detections on ridges, edges and corners[7]; it also reduces the number of detected pixels transferred to, and considered by, downstream processing stages.

In all filters discussed in Sections 3 and 4, the $\sigma$ parameter (in pixel units) is used to determine scale of analysis. It sets the extent of the impulse response (in the $n$-domain) and the bandwidth (in the $\omega$-domain) of the filter. The exact relationship between $\sigma$ and scale ($\lambda$) depends on the nature of the application and the algorithm in which the filters are used. For blob-detection via Hessian determinants, using $\lambda = 2\sigma$ was found to be appropriate.

The behavior of five bpfs (Filters A-D) in this blob-detection role was examined using real and simulated data. A two-stage lpf/hpf configuration was used in all cases for efficiency. All bpfs used the same FIR hpf stage. The $b^{\mathrm{hpf}}$ coefficients of the high-pass non-recursive filters $\mathcal{H}_d^{\mathrm{hpf}}(z)$ are $[0,1,0]$, $[1/2\,,0,-1/2]$ and $[1,-2,1]$, for $d = 0$, $1$ and $2$, respectively. The following lpf stages were used: A) FIR Gaussian, B) $4^{\mathrm{th}}$-order IIR Gaussian[10], C) FIR colored SG, D) IIR repeated pole, E) IIR Butterworth.

For $L_D = 1$ & $\bar{L}^{\mathrm{pi}} = 2$ (such that $D = 3$ & $K^{\mathrm{lpf}} = 3$), the blur stage of Filter D yields the DE coefficients in TABLE IV. as a function of pole position $p$, which determines the scale via $p = e^{-1/\sigma}$.

TABLE IV.     DE OF IIR LPF WITH REPEATED POLES AT $p = e^{-1/\sigma}$ (FILTER D)

| | $m = 0$ | $m = 1$ | $m = 2$ | $m = 3$ |
|---|---|---|---|---|
| $b_m^\circ$ | $p^3/16$ $-9p/16 + 1/2$ | — | — | — |
| $b_m^+$ | $0$ | $3\,p^4/32 - 3\,p^2/8$ $+ 9/32$ | $-3\,p^5/32 + 9\,p^3/16$ $- 15\,p/32$ | $p^6/32 - 9\,p^4/32$ $+ 9p^2/32 - 1/32$ |

| $a_m^+$ | 1 | $-3p$ | $3p^2$ | $-p^3$ |
|---|---|---|---|---|

Using $\sigma = 8$ pix for a scale of $\lambda = 16$ pix was found to be appropriate for the sample aerial imagery used in this study. Thus, the filter coefficients of the blur used in Filter D are

$b_m^+ = b_m^- = [0 \quad 0.0461 \quad -0.0773 \quad 0.0320]$

$a_m^+ = a_m^- = [1 \quad -2.6475 \quad 2.3364 \quad -0.6873]$ and

$b_m^\circ = 0.0466$.

The corresponding coefficients of the blur in Filter E designed using $\omega_c = 0.1472$ (rad/pix), i.e. $f_c = 0.0234$ (cyc/pix), are:

$b_m^+ = b_m^- = [0 \quad 0.0513 \quad -0.0420]$

$a_m^+ = a_m^- = [1 \quad -1.7929 \quad 0.8124]$ and

$b_m^\circ = 0.0518$.

The 1-D and 2-D magnitude responses of the blurs used by Filters A-E are compared in Fig. 3 and Fig. 4, respectively. The 2-D response indicates that the discrete Gaussian used in Filter A is somewhat anisotropic in the stopband. Stopband isotropy is further degraded by the recursive approximation in Filter B. The dc flatness of Filters C-E is clearly gained at the expense of isotropy in the transition band, with a "squaring" of the response along the axes around the origin. The use of recursion further degrades isotropy outside the passband, with a slower roll-off in the stopband for Filters D & E relative to Filter C. The 2-D impulse responses $h_{d_x,d_y}^{\text{bpf}}(m)$ of all filters are plotted in Fig. 5. The top row, i.e. $h_{0,0}^{\text{bpf}}(n)$, suggests that, of the flattened filters (i.e. Filters C-E) where isotropy is sacrificed for orthogonality, Filter E is the most isotropic and Filter D is the least isotropic.

Fig. 6 shows that Filter D has the desired scale selectivity (i.e. dependent on blob size) and sufficient passband isotropy (i.e. independent of blob orientation) for basic shape analysis. When the normalized Hessian is used, the centers of all ellipses, with a semi-major axis length that matches the filter scale are detected and all other ellipses ignored for a constant scale-invariant threshold (Threshold #1). Fig. 7 shows that increasing the eccentricity of the ellipses, by decreasing their width, causes spurious detections to form on the tips of all large blobs when only Threshold #1 is used. Using Threshold #2 in conjunction with Threshold #1 suppresses these detections, leaving only a single detected pixel at the centroid of the blobs with a size that is matched to the scale of the filter (see inset). The most anisotropic filter was used to create the outputs in Fig. 6 and Fig. 7 to demonstrate that the isotropy of the proposed non-Gaussian filters (Filters C-E) is sufficient in this context. All filters yielded similar detections for these synthetic inputs; although the determinant images do differ somewhat and different thresholds are required. The effect that filter anisotropy has on the Hessian determinant is apparent in Fig. 8. These images clearly show that the responses of the Gaussian filters (Filters A & B) are not perfectly isotropic and suggest that Filter C is the least anisotropic of all filters in this context, despite it being non-Gaussian.

This blob detector was also used to detect cars in real aerial imagery. One of the test images and the outputs of Filter A & D are shown in Fig. 9. As for synthetic data, all filters could be tuned for very similar behavior, which suggests that extra measures to ensure non-vanishing moments are unimportant in this context. This is perhaps unsurprising given that there is already no coupling between the $d = 1$ and $d = 2$ terms in the Hessian matrix due to symmetry. Bias errors caused by non-vanishing moments are more likely to be significant for high-order polynomial models or in coding/compression algorithms where perfect reconstruction is required. The parameters of high-order surfaces are easy to generate; though, they are difficult to interpret[29], as the structure and properties of the Hessian matrix, with two spatial derivatives in two spatial dimensions, are unique. Furthermore, local polynomial models are not a good representation of urban scenes and not much is gained by moving beyond a piece-wise constant model with step discontinuities at object boundaries. The logic of using gradients and curvature to model blobs with a constant intensity profile is somewhat questionable. However, the synthetic data results are clearly acceptable. Therefore, the use of high-order models is difficult to justify for blob pre-screening in high-throughput video surveillance systems, particularly when so much can be done with a simple quadratic model. Although recent results indicate that quartic models may be useful optic blur compensation in Hessian-based blob detectors[42]. It is also claimed in Ref. 2 that edge detectors become more orientation selective as the model order increases.

The IIR filters were initialized using Eqn. (A20) in the Appendix, whereas the FIR filters were simply initialized using zeros. Thus, Filter A produces spurious detections around the edge of the image. In practice, the FIR output images would be cropped by a margin of $K^{\text{bpf}}$ pixels to remove startup transients. For IIR filters this cannot be done because the startup transient never completely settles, it simply fades away until its effect becomes negligible.

Both filters detect most cars, although some barely discernable cars in shadows near the bottom left corner are missed. Note that using Threshold #2 eliminates detections on the ends of the bus as intended. There are many false alarms for these threshold settings. However, the results indicate that this approach may be suitable as a pre-screening algorithm, particularly when execution speeds are considered.

Execution times of various software implementations for Filters A & D are compared in TABLE V. The 1-D recursive lpf in TABLE V. (Filter D) is realized using only 13 multiply and accumulate (MAC) operations per pixel regardless of scale, i.e. the extent of the IR. By comparison, the non-recursive Gaussian (Filter A) with $\sigma = 12$, requires 121. The achieved speedup (6.9x) for a single thread is somewhat less than the MAC-analysis suggests due to the overhead of feedback and the code required to support forward/backward filtering. For the largest scale where the speedup is greatest, the speedup decreases as the number of available threads increases. This is because all pixels within a row or column are processed sequentially (via recursion), for filters with infinite IRs; whereas, each pixel may be processed independently, for filters with finite IRs. Thus, potential for parallelism is increased, which reduces the dependence of $T_E$ on $\sigma$ in the FIR/GPU case.

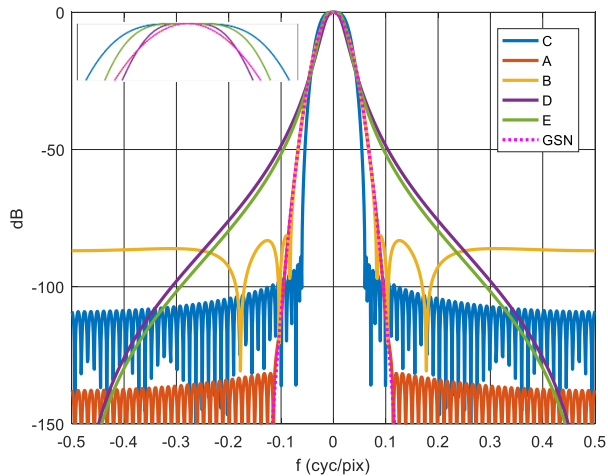

Fig. 3. 1-D magnitude response of low-pass blurs, i.e. $H_0^{\mathrm{lpf}}(\omega)$, for Filters A-E. Response of continuous Gaussian prototype also shown (GSN). Inset shows passband response over $f = \pm 0.025$ (cycles per pixel) and 0 to -3 dB.

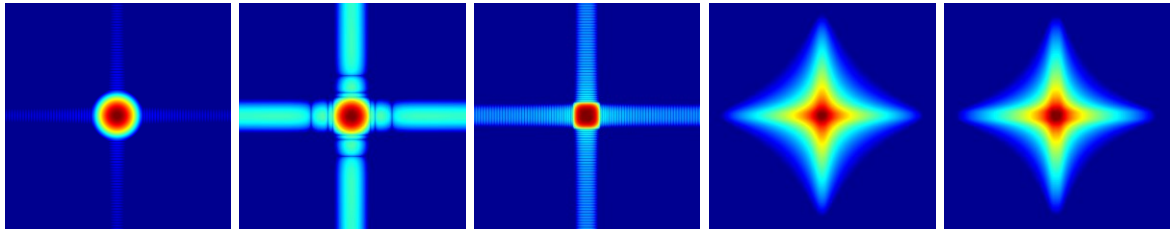

Fig. 4. 2-D magnitude response of low-pass blurs, i.e. $H_{0,0}^{\mathrm{lpf}}(\omega_x, \omega_y)$, over $\pm\pi$ and a range of 0 to -150 dB, for Filters A-E (left to right).

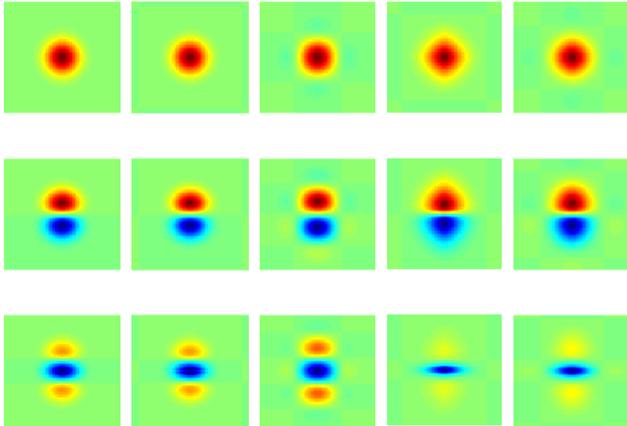

Fig. 5. The 2-D impulse responses $h_{d_x,d_y}^{\mathrm{bpf}}(m)$ of Filters A-E (left-to-right) for $d_x = 0 \ldots 2$ (top to bottom) and $d_y = 0$, truncated at $m = \pm 5\sigma$ in each dimension.

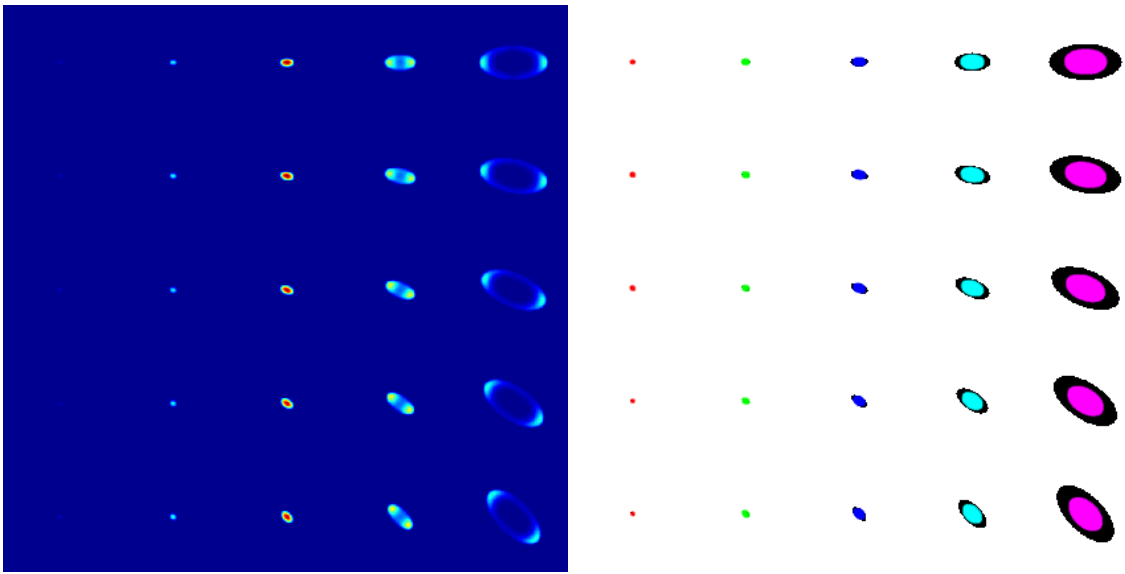

Fig. 6. Determinant of the normalized Hessian matrix for a test pattern input; processed using Filter D (see TABLE IV. ) with $\lambda = 16$, i.e. $\sigma = 8$ (left). Pixels detected using Threshold #1 for five different filters tuned using $\lambda = 4$, 8, 16, 32 & 64 pix, plotted in red, green, blue cyan and magenta, respectively, superimposed on the input (right). The input contains a series of solid black ellipses, all with an eccentricity of 2, on a white background. The semi-major axes of the ellipses are 4, 8, 16, 32 & 64 pix, from left to right. The axis is rotated from 0° through to 45° in equal increments, from top to bottom.

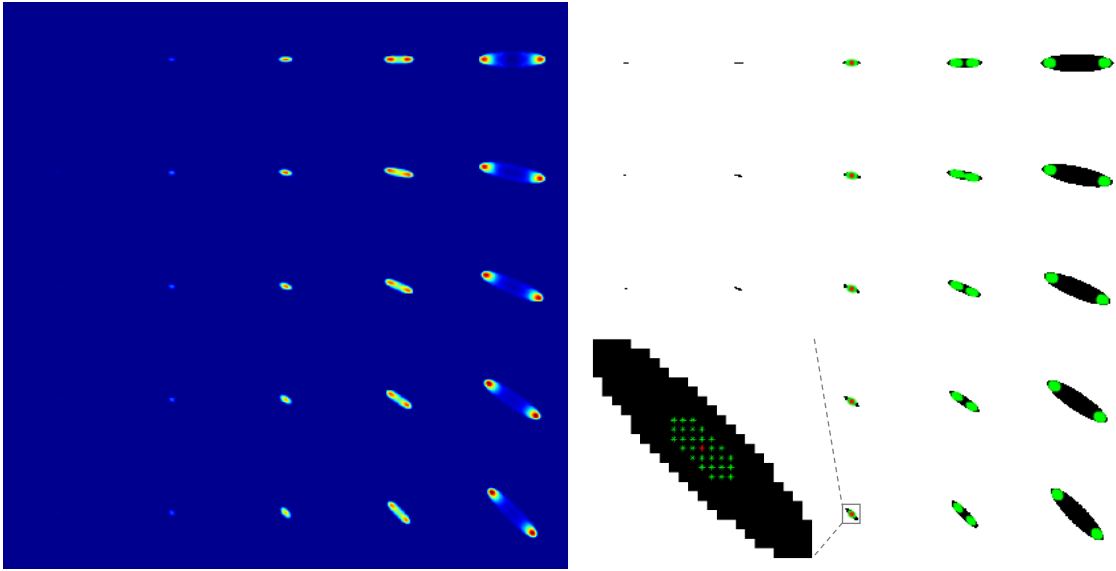

Fig. 7. Outputs for input image processed using Filter D (see TABLE IV. ) with $\lambda = 16$ (left and right). Determinant of the normalized Hessian matrix for a test pattern input (left). Pixels detected using Threshold #1 only (green) and Thresholds #1 & #2 combined (red) on the input (right). The semi-minor axes of the ellipses are now halved for an eccentricity of 4.

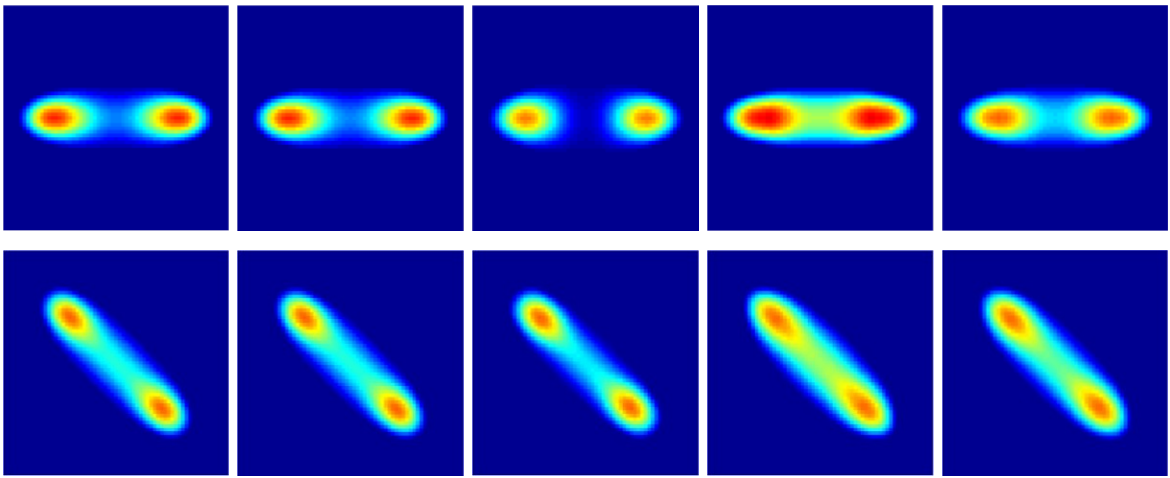

Fig. 8. Hessian determinant for Filters A-E (left to right) with $\lambda = 16$. Detail of output image around blobs with a semi-major axis of 16 pix and an eccentricity of 4 are shown, for blobs orientations 0° (top) and 45° (bottom).

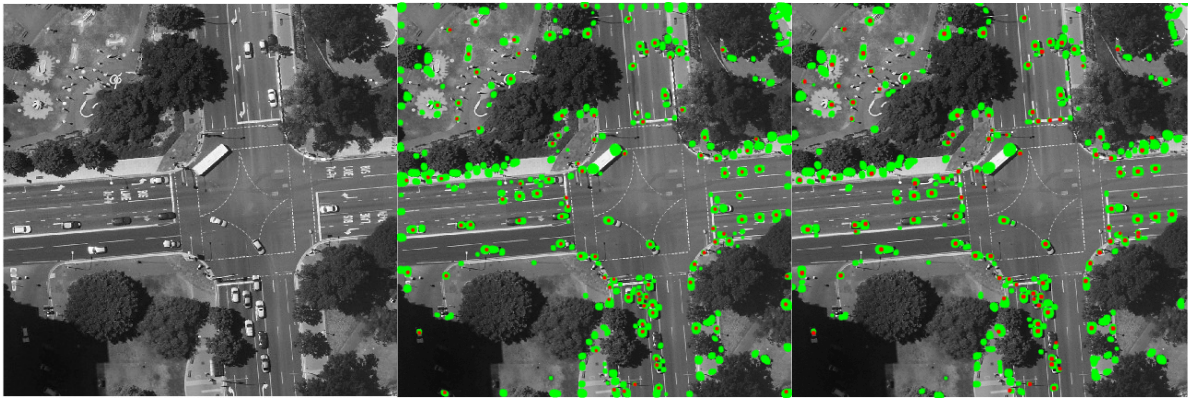

Fig. 9. Aerial image input (left) processed using filters tuned to a scale of $\lambda = 16$; Filter A (center) and Filter D (right). Detections produced using Threshold #1 (green). Detections produced using Threshold #1 and Threshold #2 (red). Processing with Filter D using a MATLAB® script (with no optional toolboxes) takes less than 0.2 s for this 1024x1024 image. See TABLE V. for CPU specifications.

TABLE V.     EXECUTION TIME ($T_E$)[*] OF C++ IMPLEMENTATIONS: NON-RECURSIVE FIR (FILTER A) VS. RECURSIVE IIR (FILTER D) REALIZATIONS[†]

| | Config. | Stage 1 (lpf) | | | | Stage 2[c] |
|---|---|---|---|---|---|---|
| $\sigma$ (pix) | | 3 | 6 | 12 | 24 | - |
| CPU[a]<br>Single thread | A | 1.70 | 3.00 | 5.80 | 11.0 | 2.2 |
| | D | 0.86 | 0.83 | 0.84 | 0.84 | |
| CPU[a]<br>Multiple threads | A | 0.47 | 0.70 | 1.30 | 2.40 | 1.1 |
| | D | 0.42 | 0.41 | 0.42 | 0.41 | |
| GPU[b]<br>Multiple threads | A | 0.30 | 0.33 | 0.42 | 0.56 | 0.11 |
| | D | 0.14 | 0.14 | 0.14 | 0.13 | |

[*] Times (s) for a single test-run. 4096 x 4096 input image.
[†] Filter A used $K^{\text{lpf}} = 5\sigma$, see Eqn. (9) for DE coefficients.
[†] Filter D used $K^{\text{lpf}} = 3$, see TABLE IV. for DE coefficients; both Configs. used $D = 3$.
[a] Intel® Core™ i7-6700HQ central processing unit (CPU) @2.6 GHz, 4 cores for a maximum of 8 concurrent threads.
[b] NVIDIA® GeForce® GTX 970M graphics processing unit (GPU), programmed using CUDA, 1280 cores.
[c] hpf application plus computation and thresholding of Hessian determinant.

## 6    Conclusion

Banks of digital filters require orthogonalization when they are used to perform shape analysis using derivatives. This process optimizes the response for the assumed polynomial model in the low-frequency passband. Without orthogonalization, for vanishing moments, filters designed using Gaussian prototypes trade passband perfection for improved transition band and stopband isotropy – so that modelling errors are smeared uniformly without creating false visual artifacts.

Filter design and tuning is an exercise in balance and compromise. Working with image filters is further complicated by the fact that performance is difficult to quantify and often assessed visually. The following design consideration is examined in this paper: Is it better to have perfect polynomial reconstruction along the orthogonal $x$- and $y$-axes and a less isotropic response, or to have reasonable polynomial reconstruction in all directions and a more isotropic response? The short answer to this question is: "it depends". Novel ways of designing FIR and IIR filters that may be configured to strike a suitable balance have been proposed.

Recursive filters are typically faster than non-recursive filters, although the speedup depends on the scale of analysis and the degree of parallelization. Of the non-Gaussian IIR filters proposed here, the discrete Butterworth filter (Filter E) has slightly improved isotropy and a slightly lower computational complexity than the repeated-pole filter (Filter D). Otherwise they are both quite similar. However, the closed-form expressions for the coefficients of the repeated-pole filter as a function of scale makes it much easier to use, particularly when online tuning is required. Savitzky-Golay filters are widely used in 1-D applications due to their polynomial interpretation in the $n$-domain and their dc flatness in the $\omega$-domain. The colored variant proposed here (Filter C) provides a simple way of adapting them to 2-D applications by improving isotropy and by allowing scale to be controlled. In curvature-based blob-detection applications Gaussian filters are sufficient. FIR and IIR realizations should be used for fine and coarse scales, respectively, for computational efficiency.

## Appendix: An introduction to recursive filtering for image processing

Most filters used in image processing have an FIR, which may be designed, analyzed and realized with a minimum of signals-and-systems theory. Furthermore, those filters are usually designed using Gaussians. Therefore, a common analytical framework that accommodates FIR and IIR filtering (see TABLE VI. ) is provided in this Appendix.

TABLE VI.    COMPARISON OF FIR AND IIR FILTER PROPERTIES

|  | Digital FIR filters | Digital IIR filters |
|---|---|---|
| Impulse response: | Truncated tails. Body may be shaped arbitrarily. Filter initialization is optional (not usually done, pad with zeros instead). | Exponentially decaying tails. Body is restricted to forms that may be generated recursively, via feedback. Must be initialized properly to avoid start-up transients. |
| Magnitude response: | Sidelobes are unavoidable but may be lowered when long IRs are used. | Highly frequency selective when long IRs are used. |
| Phase response: | Perfect phase linearity for symmetric and anti-symmetric IRs in both casual and non-casual systems. | Perfect symmetry is not possible in causal systems (with one-sided IRs). But it *is* possible in non-causal systems (with a two-sided IRs). |
| Computational complexity: | (non-recursive) Depends on the length of the IR only, not on the model order. | (recursive) Depends on the model order only, not on the duration of the IR. |

Like Deriche's formulations[10,15], IRs are expressed and realized as the *sum* of forward and backward parts in this paper. Expressing the filter's TF as the *product* of forward and backward parts via a spectral factorization, and realizing the filter by connecting the output of the first part to the input to the second part is also popular[12-14]. This approach however, is not adopted here because analyzing the sum of forward and backward responses is simpler than analyzing their product, particularly when their FRs are expressed as a linear combination of basis functions. The sum-of-parts approach is modified here by introducing a third (central) term, which further simplifies design and analysis by preserving symmetry. Notions of causality (including non-causality and anti-causality) that are routinely used in the digital signal processing literature are somewhat misleading in digital image processing because only spatial coordinates are involved where the pixels are simultaneously available. Thus, the concept of sequential time-ordering is irrelevant. Regardless, the (non-causal) 1-D digital filters used here are defined, characterized, and realized via the following:

*Definitions*

Real *impulse response (IR)* $h_d(m)$ in the $n$-domain, i.e. the discretized spatial or pixel domain, digitized using a uniform (dimensionless) sampling period ($T_s = 1$). The IR is either symmetric or antisymmetric around $m = 0$ (for a perfectly linear phase response) and defined over $m = -\infty \ldots \infty$, for IRs of infinite duration; or over $m = -K \ldots K$, for IRs of finite duration with an odd filter length of $M = 2K + 1$ pixels. The IR index $m$ and the data index $n$ (for $n = 0 \ldots N - 1$) are both in the pixel domain but increase in opposite directions. This convention is used here for non-causal 2-D systems to maintain consistency with causal 1-D terminology, where digital systems are realized using delays. In FIR and IIR cases the IR is obtained by taking the inverse $\mathcal{Z}$-transform of the corresponding TF. In the (non-recursive) FIR case it is also obtained by taking the inverse discrete-time Fourier- ($\mathcal{F}$-) transform of the corresponding FR. In the (recursive) IIR case it is usually easier to set the IR equal to the output of a DE implementation of the TF when processing a unit impulse input, until the output has effectively converged on zero.

Complex (discrete-time) *transfer function (TF)* $\mathcal{H}(z)$ in the complex $z$-domain, where $z$ and $z^{-1}$ are the unit shift operators in the forward and backward directions, respectively, denoted using the $[\cdot]^+$ & $[\cdot]^-$ superscripts. $\mathcal{H}(z)$ is determined by taking the two-sided $\mathcal{Z}$-transform of the IR using $\mathcal{H}(z) = \mathcal{Z}\{h(m)\}$, where $h(m)$ is a continuous interpolating function in the IIR case, and a linear combination of shifted unit impulses in the finite IR case. $\mathcal{H}(z)$ is a rational function of $2K$th order, i.e.

$$\mathcal{H}(z) = \mathcal{B}(z)/\mathcal{A}(z) = \sum_{m=0}^{M-1} b_{m-K} z^m / \sum_{m=0}^{M-1} a_{m-K} z^m \qquad \text{(A1a)}$$

with the roots of the $\mathcal{B}$ and $\mathcal{A}$ polynomials defining the zeros and poles of the TF, respectively. Multiplying the numerator and denominator by $z^{-K}$ yields the desired form

$$\mathcal{H}(z) = \sum_{m=-K}^{K} b_m z^m / \sum_{m=-K}^{K} a_m z^m \qquad \text{(A1b)}$$

that is centered on the zero shift ($m = 0$). FIR systems are a special limiting case with all $K$ poles at the origin of the complex $z$ plane. Thus $\mathcal{H}(z) = \sum_{m=-K}^{K} b_m z^m$, where $b_m = h_d(m)$ and $a_m = 0$ for non-zero $m$. Note that the linear combination of mixed negative and positive powers of $z$ is referred to here as a "polynomial" even though this term is usually reserved for non-negative powers. Both forms are, however, interchangeable using a shift of $K$ pixels, i.e. by multiplying the numerator and denominator polynomials by $z^K$.

Complex *frequency-response (FR)* $H(\omega)$ in the $\omega$-domain, where $\omega$ is the angular frequency in units of radians per pixel with $\omega$ real and defined over $\pm\pi$. $H(\omega)$ is determined by evaluating $\mathcal{H}(z)$ around the unit circle, i.e. by substituting $e^{i\omega}$ for $z$ in $\mathcal{H}(z)$, where $i = \sqrt{-1}$, or by taking the discrete-time $\mathcal{F}$-transform of $h(m)$, i.e.

$$H(\omega) = \mathcal{F}\{h(m)\} = \sum_{m=-K}^{K} h(m) e^{-im\omega} \tag{A2}$$

for IRs of finite duration. The inverse discrete-time $\mathcal{F}$-transform of $H(\omega)$ may then be used to interpolate the IR at non-integer $m$, i.e.

$$h(m) = \frac{1}{2\pi} \int_{\omega=-\pi}^{\pi} H(\omega) e^{im\omega} d\omega. \tag{A3}$$

Each filter is realized in the $n$-domain using a linear *difference equation (DE)*. The TF is structured in a way that yields a symmetric/antisymmetric IR that is centered on $m = 0$ in the $n$-domain, for a real/complex FR and a zero/constant phase response in the $\omega$-domain – for symmetric/anti-symmetric responses, respectively. In both FIR and IIR cases, the DE is simply obtained from the polynomial coefficients of the TF. In the (non-recursive) FIR case the DE is equal to the IR; thus, it may also be found by taking the inverse $\mathcal{Z}$-transform of the TF or the inverse discrete-time $\mathcal{F}$-transform of the FR.

*Realization*

The TF is an indispensable representation of a digital filter because it links the IR, the FR, and the DE, thus allowing the filter designer to move freely from one system representation to another [31]. In the (non-recursive) FIR case, the TF representation is an optional formality and not explicitly required, as only the discrete-time $\mathcal{F}$-transform and its inverse are required to move between the $n$- and $\omega$-domains and the DE is simply equal to the IR. However, in the (recursive) IIR case, the $z$-domain (thus the $\mathcal{Z}$-transform and its inverse) is required to connect the $n$- and $\omega$-domains and to provide the DE coefficients.

The connection between the TF and its DE for both FIR and IIR filters will now be described. Like tracking and control systems, digital filter design in image-processing systems is slightly different to digital filter design in audio and radio-frequency systems because the IR and the FR are of equal importance; furthermore, relatively compact IRs are required for a short transient response (i.e. to a step or ramp discontinuity, e.g. an object edge in an image), thus the filter's ability to resolve and discriminate frequencies is rather limited. The worked examples provided below were chosen for their ability to accommodate these somewhat unusual priorities.

In the IIR case where feedback is utilized, the recursive TF in Eq. (A1b) is not realizable because the current filter output is computed using the preceding and following filter outputs as inputs. As the DE recursion must proceed in only direction, one of these inputs is not yet available; therefore, a procedure to factor $\mathcal{H}(z)$, into a form that gives rise to a realizable DE, is required.

Non-recursive FIR filters are realized in the $n$-domain via the convolution

$$\mathcal{y}(n) = \sum_{m=0}^{M-1} b_{m-K} x(n-m) \tag{A4}$$

where $x$ and $\mathcal{y}$ (script typeface, to distinguish these variables from the spatial coordinates) are the input and output of the filter, respectively. This operation is then followed by a forward shift of $K$ pixels to compensate for the backward shift applied by the convolution. Alternatively, the convolution is applied as a single stage:

$$y(n) = \sum_{m=-K}^{K} b_m x(n-m) \tag{A5}$$

or re-written in the three-term form preferred here

$$y(n) = y^+(n) + y^-(n) + y^\circ(n) \ . \tag{A6}$$

The three terms are defined as follows:

$$y^+(n) = \sum_{m=1}^{K} b_m^+ x(n-m) \tag{A7}$$

applies the positive (causal) part of the IR, in the forward/downward direction, i.e. left-to-right or top-to-bottom, for $n = 0 \dots N-1$.

$$y^-(n) = \sum_{m=1}^{K} b_m^- x(n+m) \tag{A8}$$

applies the negative (anti-causal) part of the IR, in the backward/upward direction, i.e. right-to-left or bottom-to-top for $n = N-1 \dots 0$, assuming the image origin is at the top-left corner. For $1 \leq m \leq K$: $b_m^- = b_m^+$ for even/symmetric IRs, $b_m^- = -b_m^+$ for odd/antisymmetric IRs.

$$y^\circ(n) = b_0^\circ x(n) \tag{A9}$$

applies the central part of the IR (where $m = 0$) with $|b_0^\circ| > 0$ for even IRs and $b_0^\circ = 0$ for odd IRs.

This three-term DE is more useful when dealing with recursive IIR filters, realized in the $n$-domain via feedback. In this case, the three terms in Eq. (A6) are now defined using the more general form:

$$y^+(n) = \sum_{m=1}^{K} b_m^+ x(n-m) - \sum_{m=1}^{K} a_m^+ y^+(n-m) \tag{A10a}$$

$$y^-(n) = \sum_{m=1}^{K} b_m^- x(n+m) - \sum_{m=1}^{K} a_m^- y^-(n+m) \ \text{and} \tag{A10b}$$

$$y^\circ(n) = b_0^\circ x(n) \tag{A10c}$$

assuming $a_m^- = a_m^+$, by symmetry.

For causal 1-D systems, the filter coefficients ($b$ & $a$) of the DE are simply equated to the corresponding polynomial coefficients $\mathcal{B}(z)$ & $\mathcal{A}(z)$ of the TF, where $\mathcal{H}(z) = \mathcal{B}(z)/\mathcal{A}(z)$. For non-causal 1-D systems, the filter coefficients ($b^+, b^\circ, b^-, a^+$ & $a^-$) of the DE are also derived from the TF; however, $\mathcal{H}(z)$ must first be rearranged into a form that matches Eq. (A10). Its denominator is factored into forward and backward parts $\mathcal{A}^+(z)$ & $\mathcal{A}^-(z)$, with poles inside and outside the unit circle (respectively), with a constant term $\mathcal{A}^\circ$, i.e.

$$\mathcal{H}(z) = \mathcal{B}(z)/\mathcal{A}(z) = \mathcal{B}(z)/\mathcal{A}^\circ \mathcal{A}^+(z) \mathcal{A}^-(z), \text{ where} \tag{A11}$$

$$\mathcal{A}(z) = \mathcal{A}^\circ \mathcal{A}^+(z) \mathcal{A}^-(z) . \tag{A12}$$

We now seek an equivalent three-part form, such that

$$\mathcal{H}(z) = \mathcal{B}^+(z)/\mathcal{A}^+(z) + \mathcal{B}^\circ/\mathcal{A}^\circ + \mathcal{B}^-(z)/\mathcal{A}^-(z), \text{ i.e.} \tag{A13a}$$

$$\mathcal{H}(z) = \frac{b_{-K}z^{-K} + b_{1-K}z^{1-K} \ldots + b_{-1}z^{-1} + b_0 + b_1 z^1 \ldots + b_{K-1}z^{K-1} + b_K z^K}{a_{-K}z^{-K} + a_{1-K}z^{1-K} \ldots + a_{-1}z^{-1} + a_0 + a_1 z^1 \ldots + a_{K-1}z^{K-1} + a_K z^K} =$$

$$\frac{b_K^+ z^{-K} + b_{K-1}^+ z^{1-K} \ldots + b_1^+ z^{-1}}{a_K^+ z^{-K} + a_{K-1}^+ z^{1-K} \ldots + a_1^+ z^{-1} + 1} + \frac{b_0^\circ}{a_0^\circ} + \frac{b_1^- z^1 \ldots + b_{K-1}^- z^{K-1} + b_K^- z^K}{1 + a_1^- z^1 \ldots + a_{K-1}^- z^{K-1} + a_K^- z^K} . \tag{A13b}$$

Proceeding with the summation by placing the three terms over a common denominator

$$\mathcal{H}(z) = \frac{\mathcal{A}^\circ \mathcal{B}^+(z) \mathcal{A}^-(z) + \mathcal{B}^\circ \mathcal{A}^+(z) \mathcal{A}^-(z) + \mathcal{A}^\circ \mathcal{B}^-(z) \mathcal{A}^+(z)}{\mathcal{A}^\circ \mathcal{A}^+(z) \mathcal{A}^-(z)} \tag{A14}$$

and equating numerators, reveals that

$$\mathcal{B}(z) = \mathcal{A}^\circ \mathcal{B}^+(z) \mathcal{A}^-(z) + \mathcal{B}^\circ \mathcal{A}^+(z) \mathcal{A}^-(z) + \mathcal{A}^\circ \mathcal{B}^-(z) \mathcal{A}^+(z) \tag{A15}$$

where $\mathcal{B}^\circ = b_0^\circ$, $\mathcal{B}^+(z) = \sum_{m=1}^K b_m^+ z^{-m}$, $\mathcal{B}^-(z) = \sum_{m=1}^K b_m^- z^m$, $\mathcal{A}^\circ = a_0^\circ$, $\mathcal{A}^+(z) = 1 + \sum_{m=1}^K a_m^+ z^{-m}$ and $\mathcal{A}^-(z) = 1 + \sum_{m=1}^K a_m^- z^m$ The unknown polynomials $\mathcal{B}^+(z)$ and $\mathcal{B}^-(z)$ and the coefficient $\mathcal{B}^\circ = b_0^\circ$, of the decomposed TF are found by equating coefficients using the system of linear equations $\boldsymbol{b} = \widetilde{\mathcal{A}}\widetilde{\boldsymbol{b}}$, then solving for the numerator coefficients of the filter $\widetilde{\boldsymbol{b}} = \widetilde{\mathcal{A}}^{-1}\boldsymbol{b}$. The structure of $\widetilde{\mathcal{A}}$ is best revealed using an example; for $K = 3$:

$$\mathcal{H}(z) = \frac{b_{-3}z^{-3} + b_{-2}z^{-2} + b_{-1}z^{-1} + b_0 + b_1z^1 + b_2z^2 + b_3z^3}{a_{-3}z^{-3} + a_{-2}z^{-2} + a_{-1}z^{-1} + a_0 + a_1z^1 + a_2z^2 + a_3z^3}$$

$$= \frac{b_3^+z^{-3} + b_2^+z^{-2} + b_1^+z^{-1}}{a_3^+z^{-3} + a_2^+z^{-2} + a_1^+z^{-1} + 1} + \frac{b_0^\circ}{a_0^\circ} + \frac{b_1^-z^1 + b_2^-z^2 + b_3^-z^3}{1 + a_1^-z^1 + a_2^-z^2 + a_3^-z^3} \tag{A16}$$

thus like-terms are matched using

$$\boldsymbol{b} = \begin{bmatrix} b_{-3} \\ b_{-2} \\ b_{-1} \\ b_0 \\ b_1 \\ b_2 \\ b_3 \end{bmatrix}, \widetilde{\boldsymbol{\mathcal{A}}} = \begin{bmatrix} 1 & 0 & 0 & a_{-3}^\pm & 0 & 0 & 0 \\ a_1^- & 1 & 0 & a_{-2}^\pm & a_3^+ & 0 & 0 \\ a_2^- & a_1^- & 1 & a_{-1}^\pm & a_2^+ & a_3^+ & 0 \\ a_3^- & a_2^- & a_1^- & a_0^\pm & a_1^+ & a_2^+ & a_3^+ \\ 0 & a_3^- & a_2^- & a_1^\pm & 1 & a_1^+ & a_2^+ \\ 0 & 0 & a_3^- & a_2^\pm & 0 & 1 & a_1^+ \\ 0 & 0 & 0 & a_3^\pm & 0 & 0 & 1 \end{bmatrix}, \widetilde{\boldsymbol{b}} = \begin{bmatrix} a_0^\circ b_3^+ \\ a_0^\circ b_2^+ \\ a_0^\circ b_1^+ \\ b_0^\circ \\ a_0^\circ b_1^- \\ a_0^\circ b_2^- \\ a_0^\circ b_3^- \end{bmatrix}. \tag{A17}$$

The $a_m^\pm$ coefficients are taken from the polynomial $\mathcal{A}^\pm(z)$, which is formed from the product of the denominator polynomials of the forward and backward filters

$$\mathcal{A}^\pm(z) = \sum_{m=-K}^{K} a_m^\pm z^m = \mathcal{A}^+(z)\mathcal{A}^-(z) = (1 + \sum_{m=1}^{K} a_m^+ z^{-m})(1 + \sum_{m=1}^{K} a_m^- z^m). \tag{A18}$$

The forward and backward coefficients are determined using the $K$ roots $(p_m)$ of $\mathcal{A}^+(z)$

$$\mathcal{A}^+(z) = (1 + \sum_{m=1}^{K} a_m^+ z^{-m}) = z^{-K} \prod_{m=1}^{K}(z - p_m) \tag{A19}$$

for real or complex $p_m$ with $|p_m| < 1$. For symmetric and antisymmetric responses, the backward coefficients are then simply set using $a_m^- = a_m^+$. As $\mathcal{A}^+(z)$ and $\mathcal{A}^-(z)$ are formed from products in a manner that yields $a_0^+ = a_0^- = 1$, the $a_0^\circ$ coefficient ensures that the factored denominator product, i.e. $\mathcal{A}^\circ \mathcal{A}^+(z)\mathcal{A}^-(z)$, is normalized properly; it is set using $\mathcal{A}^\circ = a_0^\circ = a_0/a_0^\pm$.

This decomposition, made possible by the three-term representation of the IR, is primarily intended for recursive IIR filters. It is however, sufficiently general to accommodate non-recursive FIR filters, with $\widetilde{\boldsymbol{\mathcal{A}}}$ being the identity matrix $\boldsymbol{I}$. It provides a simple way of converting a non-realizable non-causal IIR filter into a realizable non-causal IIR filter.

The design procedure described above allows a non-realizable TF to be defined in the $z$-domain, that approximately satisfies a set of requirements in the $n$- and $\omega$-domains. The realizable TF is then reached via the factorization in Eqn. (A16). In the worked examples that follow, the coefficients of simple low-pass filters are derived to illustrate the decomposition described above. In the first example, $\mathcal{H}(z)$ is derived from qualitative IR and FR considerations via consecutive convolutions with (two-sided) exponential pulses; in the second example, a Butterworth analog prototype is discretized using the bilinear transformation.

*Example 1: Convolved (two-sided) exponential pulses*

The use of a 1-D sliding rectangular (i.e. finite and uniformly weighted) window is a reasonable way of computing a local mean or the magnitude of any dc component in some applications; furthermore, it may be applied recursively, for a low-pass filter with an FIR[7-9]. As the length ($M$) of the window decreases, the spatial selectivity increases and the frequency selectivity of the filter decreases, so its response to other near-dc frequencies increases. The FR of this FIR filter is the Dirichlet kernel[Error! Reference source not found.]. The Dirichlet kernel (also known as the periodic sinc function) embodies the shortcomings of using rectangular windows for the estimation of dc components in 1-D: It has unity gain at dc (essential) and zero gain where the window length is a whole multiple of the wavelength (desirable) but it also has a non-negligible gain at the intervening frequencies (undesirable). Replacing the hard truncation of the window with a smooth taper in the pixel domain reduces the height of side-lobes in the frequency domain by eliminating the conditions that give rise to the pronounced anti-resonance phenomena. A non-recursive FIR window can of course be tapered arbitrarily to achieve the desired balance between main-lobe width and side-lobe height.

Reasonable recursive filters with a low-pass FIR may be reached by convolving multiple rectangular pulses, where each pulse is formed by subtracting the output of a delayed accumulator from the output of a current accumulator[7-9]. With each convolution, the resulting response (in both the $n$- and $\omega$-domains) becomes progressively more circularly symmetric or isotropic in 2-D with thinner tails in 1-D. When each accumulator is implemented recursively using a marginally stable pole at $z = 1$ (cancelled by a zero) i.e. the number of multiply and add operations is independent of the impulse response duration; however, the number of shift operations and registers is approximately equal to $M$.

Alternatively, convolving multiple (two-sided) exponential pulses, where the sides of each pulse are formed using so-called "leaky" accumulators[47], each with a pole on the positive real axis at $z = p$ and $z = 1/p$, yields the recursive non-causal low-pass filter with an IIR and the following TF:

$$\mathcal{H}(z) = c_0 \left\{ \frac{1}{(z^{-1}-p)(z-p)} \right\}^K = c_0 \left\{ \frac{1}{-pz^{-1}+(1+p^2)-pz} \right\}^K \tag{A20}$$

where $K$ is the number of pulses and $c_0$ is a normalizing factor which ensures that $H(\omega)|_{\omega=0}$, i.e. $\mathcal{H}(z)|_{z=1}$, is equal to unity. The impulse response of each (normalized) exponential pulse is $h(m) = e^{-|m|/\lambda}$ or $h(m) = p^{|m|}$ where $p = e^{-1/\lambda}$. Thus, the nominal "scale" of the pulse, i.e. $\lambda$ (in pixels), is encoded in $p$.

The 1-D IR of the two-sided exponential used to make the TF in Eq. (A20) has a cusp-like discontinuity at $m = 0$. The sharpness of the cusp may however be reduced if high frequencies are attenuated by placing zeros at $z = 1$ for a Nyquist null (i.e. at $\omega = \pi$). This yields a more rounded and isotropic response in 2-D, near the origin (in both the $n$- and $\omega$-domains). The TF of the low-pass filter built from these "blunt" exponentials[47] is

$$\mathcal{H}(z) = c_0 \left\{ \frac{z^{-1} + 2 + z}{-pz^{-1} + (1+p^2) - pz} \right\}^K. \tag{A21}$$

For both low-pass IIR filter variants, the number of arithmetic *and* shift operations is independent of the IR duration.

The two-sided exponential pulse is first-order in each direction and it is the simplest form of recursive taper. It eliminates the side-lobes of the Dirichlet kernel and replaces them with monotonic decay; For the blunt exponential, $|H(\omega)|$ decays to zero at $\omega = \pi$. The narrower main-lobe and the absence of side-lobes do not, however, make this response ideal; for instance, the gradual decay of the tails in the $n$-domain and the lack of bandwidth in the $\omega$-domain may be problematic.

For responses derived from recursively-generated FIR pulses, it is easiest to realize the filter via a series of sequential convolutions. A similar approach may also be taken for the two-sided exponential in Eq. (A20) by summing the outputs of two first-order filters applied in the forward and backward directions on each of the $K$ convolutions; however, the filtering operation for the IIR cases in Eqs. (A20) & (A21) is further simplified if it is realized via a single convolution by summing the outputs of two $K$th-order filters. The coefficients of the required $K$th-order filter (the same in both directions, due to symmetry) are found using Eq. (A17). The use of three (rectangular or exponential) pulses is generally regarded as the minimum order required for a low-pass filter with a reasonably isotropic response. Using $K = 3$ in Eq. (A21) yields

$$\mathcal{H}(z) = c_0 \frac{\mathcal{B}(z)}{\mathcal{A}(z)} \text{ with}$$

$$\mathcal{B}(z) = (z^{-1} + 2 + z)^3$$

$$= z^{-3} + 6z^{-2} + 15z^{-1} + 20 + 15z + 6z^2 + z^3$$

$$\mathcal{A}(z) = (-pz^{-1} + 1 + p^2 - pz)^3$$

$$= -p^3 z^{-3} + 3(p^4 + p^2)z^{-2} - (3p^5 + 9p^3 + 3p)z^{-1} + (p^6 + 9p^4 + 9p^2 + 1) -$$

$$(3p^5 + 9p^3 + 3p)z + 3(p^4 + p^2)z^2 - p^3 z^3 \text{ and}$$

$$c_0 = \frac{p^6}{64} + \frac{-3p^5}{32} + \frac{15p^4}{64} + \frac{-5p^3}{16} + \frac{15p^2}{64} + \frac{-3p}{32} + \frac{1}{64} . \tag{A22}$$

From Eq. (A19) we have

$$\mathcal{A}^+(z) = -p^3 z^{-3} + 3p^2 z^{-2} - 3pz^{-1} + 1 \text{ and}$$

$$\mathcal{A}^-(z) = 1 - 3pz + 3p^2 z^2 - p^3 z^3 \text{ thus}$$

$$a_1^+ = a_1^- = -3p, a_2^+ = a_2^- = 3p^2, a_3^+ = a_3^- = -p^3 . \tag{A23}$$

Note that in this example $\mathcal{A}^+(z)\mathcal{A}^-(z) = \mathcal{A}(z)$, thus $\mathcal{A}^\circ(z) = a_0^\circ = 1$ and Eq. (A17) becomes

$$\widetilde{\mathcal{A}} = \begin{bmatrix} 1 & 0 & 0 & -p^3 & 0 & 0 & 0 \\ -3p & 1 & 0 & 3(p^4+p^2) & -p^3 & 0 & 0 \\ 3p^2 & -3p & 1 & -(3p^5+9p^3+3p) & 3p^2 & -p^3 & 0 \\ -p^3 & 3p^2 & -3p & (p^6+9p^4+9p^2+1) & -3p & 3p^2 & -p^3 \\ 0 & -p^3 & 3p^2 & -(3p^5+9p^3+3p) & 1 & -3p & 3p^2 \\ 0 & 0 & -p^3 & 3(p^4+p^2) & 0 & 1 & -3p \\ 0 & 0 & 0 & -p^3 & 0 & 0 & 1 \end{bmatrix} \boldsymbol{\ell} = \begin{bmatrix} 1 \\ 6 \\ 15 \\ 20 \\ 15 \\ 6 \\ 1 \end{bmatrix} . \tag{A24}$$

After solving for forward/backward/central numerator coefficients $b$ and multiplying by the normalizing factor $c_0$ we obtain

$$b_0^\circ = \frac{-p^3}{32} + \frac{3p^2}{16} + \frac{-15p}{32} + \frac{5}{16}$$

$$b_1^+ = b_1^- = \frac{-3p^4}{64} + \frac{3p^3}{16} + \frac{-3p^2}{16} + \frac{-3p}{16} + \frac{15}{64}$$

$$b_2^+ = b_2^+ = \frac{3p^5}{64} + \frac{-9p^4}{32} + \frac{15p^3}{32} + \frac{3p^2}{16} + \frac{-33p}{64} + \frac{3}{32} \text{ and}$$

$$b_3^+ = b_3^- = \frac{-p^6}{64} + \frac{3p^5}{32} + \frac{-15p^4}{64} + \frac{15p^2}{64} + \frac{-3p}{32} + \frac{1}{64} . \tag{A25}$$

*Example 2: Butterworth filter*

Low-pass recursive (FIR or IIR) filtering in 2-D with 1-D pulses – rectangular, two-sided exponential or blunt exponential – is a straightforward/reasonable approach and the response becomes smoother and more isotropic with just a few additional passes. For IIR filters derived from exponentials, using repeated real poles allows expressions for the filter coefficients to be derived as a function of the nominal scale, using $p = e^{-1/\lambda}$. Using unique complex poles allows FRs with wider/flatter pass-bands and narrower transition bands to be crafted. In these cases, it is not obvious how the poles should be arranged, therefore analog prototypes may be used for guidance[1,48].

The FR of the analog Butterworth filter is maximally flat at dc, which may be used to support polynomial fitting and derivative estimation in the pixel domain, and has monotonic roll off, which inhibits ripples in the tails of the IR. Fortunately, these properties are maintained when the bilinear transform is used to discretize the continuous TF, i.e.

$$\mathcal{H}(z) = \mathcal{H}(s)\big|_{s=2\frac{z-1}{z+1}} \text{ where } \mathcal{H}(s) = 1\Big/\left\{1 + \left(\frac{-s^2}{\omega_c^2}\right)^K\right\}. \tag{A26}$$

The cut-off frequency $\omega_c$, determines the bandwidth of the low-pass filter; it set here using $\omega_c = 1/\lambda$, where $\lambda$ is again the nominal "scale" of the IR. For $K = 3$ we have

$$\mathcal{H}(z) = \mathcal{B}(z)/\mathcal{A}(z) \text{ with}$$

$$\mathcal{B}(z) = \omega_c^6(z^{-3} + 6z^{-2} + 15z^{-1} + 20 + 15z + 6z^2 + z^3) \text{ and}$$

$$\mathcal{A}(z) = (\omega_c^6 - 64)z^{-3} + (6\omega_c^6 + 384)z^{-2} + (15\omega_c^6 - 960)z^{-1} + (20\omega_c^6 + 1280) +$$

$$(15\omega_c^6 - 960)z + (6\omega_c^6 + 384)z^2 + (\omega_c^6 - 64)z^3. \tag{A27}$$

Thus $a_0 = 20\omega_c^6 + 1280$, i.e. the constant coefficient of $\mathcal{A}(z)$. The $b$ coefficients in the $\boldsymbol{b}$ vector of Eq. (A17) are equated to the corresponding coefficients of the $\mathcal{B}(z)$ polynomial. Solving $\mathcal{A}(z) = 0$ and using the $K$ roots inside the unit circle (i.e. where $|z| < 1$) allows $\mathcal{A}^+(z)$ to be defined as follows:

$$\mathcal{A}^+(z) = z^{-K}\left\{z - \frac{4 - \omega_c^2 + 2\sqrt{3}\omega_c i}{\omega_c^2 + 2\omega_c + 4}\right\}\left\{z - \frac{(2 - \omega_c)}{(2 + \omega_c)}\right\}\left\{z - \frac{4 - \omega_c^2 - 2\sqrt{3}\omega_c i}{\omega_c^2 + 2\omega_c + 4}\right\}. \tag{A28}$$

The $a^+$ coefficients are then equated to the coefficients of this expanded polynomial, using $\mathcal{A}^+(z) = a_3^+ z^{-3} + a_2^+ z^{-2} + a_1^+ z^{-1} + 1$, thus the $a$ coefficients of the realizable non-casual DE are

$a_1^+ = \frac{3\omega_c^3 + 4\omega_c^2 - 8\omega_c - 24}{\omega_c^3 + 4\omega_c^2 + 8\omega_c + 8}$

$a_2^+ = \frac{3\omega_c^2 - 10\omega_c + 12}{\omega_c^2 + 2\omega_c + 4}$ and

$a_3^+ = \frac{\omega_c^3 - 4\omega_c^2 + 8\omega_c - 8}{\omega_c^3 + 4\omega_c^2 + 8\omega_c + 8}$ .                              (A29)

Noting that $\mathcal{A}^-(z) = \mathcal{A}^+(z)|_{z=1/z}$ i.e. $\mathcal{A}^-(z) = 1 + a_1^- z + a_2^- z^2 + a_3^- z^3$, it follows that $a_m^- = a_1^+$. The $a^\pm$ coefficients are then set equal to the polynomial coefficients of $\mathcal{A}^\pm(z) = \mathcal{A}^+(z)\mathcal{A}^-(z) = \mathcal{A}(z)/\mathcal{A}^\circ$. From the constant coefficient of $\mathcal{A}^\pm(z)$, we have

$a_0^\pm = (20\omega_c^6 + 1280)/(\omega_c^3 + 4\omega_c^2 + 8\omega_c + 8)^2$ thus

$a_0^\circ = a_0/a_0^\pm = (\omega_c^3 + 4\omega_c^2 + 8\omega_c + 8)^2$.                              (A30)

The $\widetilde{\mathcal{A}}$ matrix may now be populated using all $a$ coefficients determined above. Solving for the $\widetilde{\boldsymbol{b}}$ vector, and dividing the forward and backward elements by $a_0^\circ$, yields the $b$ coefficients of the realizable DE

$b_0^\circ = (3\omega_c^6 + 20\omega_c^5 + 64\omega_c^4 + 120\omega_c^3 + 128\omega_c^2 + 64\omega_c)/3$

$b_1^+ = b_1^- = (4\omega_c^5 + 32\omega_c^4 + 104\omega_c^3 + 128\omega_c^2 + 64\omega_c)/3\, a_0^\circ$

$b_2^+ = b_2^- = (8\omega_c^5 + 32\omega_c^4 - 128\omega_c^2 - 128\omega_c)/3\, a_0^\circ$ and

$b_3^+ = b_3^- = (4\omega_c^5 - 8\omega_c^3 + 64\omega_c)/3\, a_0^\circ$ .                              (A31)

*IIR filter initialization*

Initialization of IIR filters in online 1-D temporal systems is optional because the startup transients become less relevant and are quickly forgotten as more data are collected; however, in 2-D spatial systems the startup transients may be difficult to ignore when images are small and the IR long. Initializing IIR filters correctly, minimizes the impact of start-up transients that would otherwise corrupt the output pixels around the perimeter of the image. This is done by assuming that initial pixel values at the edge of the image frame are held to $\pm\infty$ outside the image frame. The internal states of the filter are then set equal to the steady-state values of their respective responses to a step input with a magnitude equal to the initial input[43]. This means that the internal layout of the filter realization behind the interface must be known. For example, the $\boldsymbol{y} = \texttt{filter}(\boldsymbol{b}, \boldsymbol{a}, \boldsymbol{x}, \boldsymbol{w}_0)$ function of MATLAB® uses a transposed direct-form II realization. In this canonical form, the $\boldsymbol{b}$ coefficients are applied when the input is "distributed" over the internal states as it enters the system, which improves numerical robustness but complicates analysis somewhat. The (non-transposed) direct-form II realization, however, applies the $\boldsymbol{b}$ coefficients when the internal states are "collected" to form the output as it leaves the system, which simplifies analysis because the steady-state values of the internal states are independent of the system zeros. Let the steady-state values of the latter canonical form be $\boldsymbol{v}_0$ and let $\mathbb{T}$ be the transformation that converts the coordinate system that gives rise to the latter canonical form into one that yields the former canonical form, constructed using the observability matrices of both linear state-space systems[44,45]. Thus, the internal states of the transposed direct-form II realization for a third-order IIR filter (applied in either the forward or backward directions) are initialized using $\boldsymbol{w}_0 = \mathbb{T}\boldsymbol{v}_0$, where

$$\boldsymbol{v}_0 = \begin{bmatrix} 1 \\ 1 \\ 1 \end{bmatrix} \frac{x_0}{1 + a_1 + a_2 + a_3} \text{ and} \tag{A32a}$$

$$\mathbb{T} = \begin{bmatrix} b_1 & b_2 & b_3 \\ b_2 & b_3 + a_1 b_2 - a_2 b_1 & a_1 b_3 - a_3 b_1 \\ b_3 & a_1 b_3 - a_3 b_1 & a_2 b_3 - a_3 b_2 \end{bmatrix}. \tag{A32b}$$

Note that $\boldsymbol{b} = [0 \quad b_1 \quad b_2 \quad b_3]$ and $\boldsymbol{a} = [1 \quad a_1 \quad a_2 \quad a_3]$; thus,

in the forward direction: $\boldsymbol{b} = [0 \quad b_1^+ \quad b_2^+ \quad b_3^+]$ and $\boldsymbol{a} = [1 \quad a_1^+ \quad a_2^+ \quad a_3^+]$;

in the backward direction: $\boldsymbol{b} = [0 \quad b_1^- \quad b_2^- \quad b_3^-]$ and $\boldsymbol{a} = [1 \quad a_1^- \quad a_2^- \quad a_3^-]$.

Note also that backward filtering a row input vector $\boldsymbol{x}$ is achieved by flipping it left/right before passing it to the `filter()` function; the row output vector $\boldsymbol{y}$ is then flipped left/right to obtain the output of the backward filter. Column filtering is achieved in the same way, with up/down instead of left/right flipping operations.